\documentclass[11pt,letter]{article}

\usepackage[utf8]{inputenc}
\usepackage[T1]{fontenc}
\usepackage{lmodern}
\usepackage[margin=1in]{geometry}
\usepackage{amsmath,amssymb}
\usepackage{booktabs}
\usepackage{graphicx}
\usepackage{subcaption}
\usepackage[hidelinks]{hyperref}
\usepackage{cleveref}          
\usepackage{enumitem}
\usepackage{soul}
\usepackage[numbers,sort&compress]{natbib}
\usepackage{xcolor}
\title{Building Power Grid Models from Open Data:\\
       A Complete Pipeline from OpenStreetMap to Optimal Power Flow}
\author{Andrea Britto$^{1}$, Thiago Spina$^{1}$, Weiwei Yang$^{1}$, Spencer Fowers$^{1}$, \\Baosen Zhang$^{1,2}$ and Chris White$^{1}$\\[4pt]
        \textit{$^{1}$Microsoft Research}\\
        \textit{$^{2}$University of Washington}}
\date{May 2026}

\begin{document}
\maketitle

\begin{abstract}
Access to realistic transmission grid models is essential for power
systems research, yet detailed network data in the United States
remains restricted under critical-infrastructure regulations.
We present a pipeline that constructs complete,
OPF-solvable transmission network models entirely from publicly
available data.  The five-stage pipeline (1)~extracts power
infrastructure from OpenStreetMap via a local Overpass~API instance,
(2)~reconstructs bus-branch topology through voltage inference,
line merging, and transformer detection,
(3)~estimates electrical parameters using voltage-class lookup
tables calibrated with U.S.\ Energy Information Administration (EIA)
plant-level data,
(4)~allocates hourly demand from EIA-930 to individual buses using
US~Census population as a spatial proxy, and
(5)~solves both DC and AC optimal power flow using PowerModels.jl
with a progressive relaxation strategy that automatically loosens
constraints on imprecise models.
We validate the pipeline on all 48~contiguous US states and six
multi-state regions, including the full Western (5{,}076~buses) and Eastern
(21{,}697~buses) Interconnections.
Of the 48~single-state models, 42~(88\%) converge at the strictest
relaxation level for AC-OPF at peak hour and 44~(92\%) off-peak.
Dispatch costs (median \$22/MWh) and system losses
(median 1.0\%) are consistent with real wholesale-market outcomes.
The pipeline relies exclusively on open data sources,
enabling reproducible grid analysis without
proprietary data.  All 54~models (48~single-state and 6~multi-state) are
publicly released at
\url{https://github.com/microsoft/GridSFM}.
\end{abstract}

\section{Introduction}\label{sec:intro}


The electrical grid is under greater stress today than at any point in
its history.  Three converging forces are simultaneously reshaping demand
patterns, generation portfolios, and system failure modes:

\begin{itemize}[leftmargin=*]

\item {AI-driven demand growth.}
Large-scale data centers are now among the fastest-growing sources of
electricity demand in advanced economies.  The International Energy
Agency projects that global data-center electricity consumption will
more than double between 2024 and 2030, driven primarily by the
computational requirements of training and serving large language models
and other AI workloads~\citep{iea_energy_ai2025}.  A single hyperscale campus can
consume as much electricity as a mid-sized city, and clusters of such
facilities are increasingly concentrated in regions whose transmission
infrastructure was not designed for dense, around-the-clock industrial
loads.

\item {Renewable integration.}
Solar and wind now account for a large and growing share of new US
electricity generation, fundamentally altering how power systems
operate~\citep{eia_epa2024}.  Unlike conventional plants, renewable
resources are geographically constrained and inherently variable,
requiring detailed understanding of how they interact with dispatchable
(operator-controlled) generation during periods of stress.

\item {Extreme weather events.}
Winter Storm Uri (February 2021) triggered cascading failures across the
Texas grid, leaving 4.5~million customers without power and causing an
estimated \$195~billion in damages~\cite{utaustin2021}.  Hurricane Maria
(September 2017) caused prolonged, island-wide power outages in Puerto Rico.
Weather-related grid disruptions in the United States have approximately
doubled over the past two
decades~\cite{climatecentral2024} -- creating an urgent need for tools
that can model cascading failures and support resilient system planning.

\end{itemize}

Addressing these challenges requires quantitative models of the
transmission grid -- network representations of substations and
transmission lines with realistic electrical parameters, generator
characteristics, and demand distributions.  Such models are
essential to:

\begin{enumerate}[leftmargin=*]
\item {Predict congestion and plan transmission expansion.}
  Identifying network bottlenecks where power flows approach thermal
  limits, and evaluating where new lines or upgrades would most
  effectively relieve constraints.
\item {Optimize generator dispatch.}
  Solving the Optimal Power Flow (OPF) problem -- determining the
  least-cost combination of generators that meets demand while
  respecting all network constraints (thermal limits, voltage bounds,
  power balance).  OPF and the other power-systems concepts used in this
  paper are formally defined in \cref{sec:background}.
\item {Simulate cascading failures.}
  Modeling how the loss of a single line or generator propagates through
  the network, tripping protections and shedding load -- the mechanism
  behind large-scale blackouts.
\item {Assess resilience under stress scenarios.}
  Evaluating grid performance under extreme conditions: simultaneous
  high demand, low renewable output, and equipment outages.
\item {Enable data-driven methods.}
  Training and validating machine-learning models for congestion
  forecasting, anomaly detection, and operational decision support, all
  of which require large volumes of realistic grid states.
\end{enumerate}

In short, grid data is foundational infrastructure for the energy
transition.  Without it, researchers cannot study the systems they aim to
improve.  Yet as the next section describes, this data is largely
inaccessible.

\subsection{The Grid Data Problem}\label{sec:grid-data-problem}

The electrical transmission grid is among the most complex engineered
systems in the world, but detailed models of its topology and operating
parameters are largely inaccessible to the research community.  In the
United States, the Federal Energy Regulatory Commission (FERC)
classifies detailed grid topology, line impedance data (the electrical
characteristics of transmission lines, defined in \cref{sec:primer}),
and operating parameters as Critical Energy Infrastructure Information
(CEII)~\citep{ferc2023}.  The North American Electric Reliability
Corporation (NERC) further restricts access through its Critical
Infrastructure Protection standards, which mandate physical and cyber
security protections for bulk electric system
data~\citep{nerc2023}.  In practice, this means that the network models
used for transmission planning, reliability assessment, and market
simulation are treated as sensitive critical-infrastructure information.

For researchers, the consequences are significant.  Obtaining access to
real grid models requires a formal application to FERC, a process that
can take months and imposes strict redistribution restrictions -- a
researcher who receives CEII data cannot share it with collaborators at
other institutions, let alone publish it.  Commercial alternatives
exist, but annual license fees range from \$50,000 to \$500,000, and
the terms typically prohibit sharing derived models or results in a form
that could reconstruct the original data.

The academic community has long relied on standardized test cases to
fill this gap.  The IEEE 14-bus, 30-bus, 118-bus, and 300-bus
systems~\citep{glover2012power} -- where ``bus'' denotes a network node,
typically a substation (see \cref{sec:primer}) -- have been workhorses for power systems
research for decades.  More recently, the IEEE PES Power Grid Library
(PGLib-OPF)~\citep{pglib2019} has curated a collection of validated OPF
benchmark cases ranging from 3-bus pedagogical examples to 10{,}000+
bus networks.  However, even PGLib's largest cases are designed to test
solver algorithms, not to represent specific real-world grids.  They
predate large-scale renewable integration, lack High-Voltage Direct
Current (HVDC) links (long-distance DC corridors that connect separate
AC grids, defined in \cref{sec:primer}), and do not capture the geographic and operational diversity of
a continental-scale power system.  Birchfield
et al.~\cite{birchfield2017} showed that standard test cases fail to
reproduce key structural characteristics of real networks, including
degree distributions, electrical diameter, and generation-load spatial
patterns.

More recently, the Texas A\&M (TAMU) synthetic grid
project~\citep{xu2017irep} produced geographically realistic
2{,}000-bus and 10{,}000-bus test cases by statistically generating
topologies that match real network properties.  While valuable, these
synthetic grids are constructed to \emph{resemble} the real grid without
being \emph{derived from} it -- their geographic correspondence to actual
infrastructure is approximate, and they cannot be updated as the real
grid evolves.

This situation -- where grid data is simultaneously indispensable for
research and inaccessible to most researchers -- motivates the present
work.

\subsection{Contribution and Paper Organization}\label{sec:contribution}

This paper presents a complete pipeline that transforms raw
OpenStreetMap (OSM) data -- a collaborative, freely available geographic
database that includes mapped power infrastructure worldwide
(\cref{sec:osm}) -- into solver-ready OPF models.  The pipeline has
been validated across all 48~continental US states and multi-state
regional models up to full interconnection scale.  The goal is not to
recover the true operational grid, but to generate structurally and
electrically plausible transmission models that converge under AC-OPF,
preserve real geographic correspondence, and reproduce system-level
statistics within realistic ranges.  Every data
source used is publicly and freely accessible; the complete list of data
sources, software libraries, and their roles is given in
\cref{sec:architecture}.

The pipeline consists of five sequential stages:

\begin{enumerate}[leftmargin=*]
\item \textbf{Data Extraction} (\cref{sec:extraction}):
  Download power infrastructure from OSM.
\item \textbf{Topology Reconstruction} (\cref{sec:topology}):
  Build a network model with inferred connectivity and circuit
  classification.
\item \textbf{Parameter Estimation} (\cref{sec:parameters}):
  Assign electrical parameters and generator characteristics.
\item \textbf{Demand Allocation} (\cref{sec:demand}):
  Distribute real demand data to network buses.
\item \textbf{Optimal Power Flow} (\cref{sec:opf}):
  Solve OPF with a progressive relaxation strategy.
\end{enumerate}

\noindent
In addition to the pipeline methodology, we publicly release the
complete set of solved models -- 48~single-state and
6~multi-state regional networks -- as PowerModels-compatible JSON files
at \url{https://github.com/microsoft/GridSFM}, providing the research
community with ready-to-use OPF benchmarks derived from real
infrastructure.

The remainder of this paper is organized as follows.
\Cref{sec:background} provides essential power-systems background and
describes the OSM data source.
\Cref{sec:system} describes the pipeline architecture, data sources,
and each of the five pipeline stages in detail.
\Cref{sec:results} presents results for single-state and multi-state
models.
\Cref{sec:discussion} discusses limitations and comparisons with existing
approaches.
\Cref{sec:conclusion} concludes.
Appendices provide detailed parameter tables and solver configuration.

\section{Background}\label{sec:background}

This section defines power systems concepts and data sources used
throughout the paper. Readers familiar with power systems modeling and operations may skip to \cref{sec:system} and readers interested in more details may consult standard references~\cite{bergen2009power,glover2012power,kirschen2024power,Low_draft}. 
\subsection{Power Systems Primer}\label{sec:primer}
A power system consists of generators and loads connected by transmission lines and transformers. Their behavior is described using quantities such as voltages, currents, and power which are sinusoidal functions of time. Assuming three-phase balanced AC operations, these quantities can be described as complex numbers, also called complex phasors. Following standard electrical engineering convention, we use $j=\sqrt{-1}$ as the imaginary unit.  The system is naturally described as a graph with buses connected by branches. 

A bus is a node in the graph that represents a point where electrical equipment connects. It can represent a generator, a load, or aggregation of devices. A bus is associated with a complex voltage, and the set of bus voltages can be thought of as the state of the system. The voltage magnitudes are typically tightly controlled at specific levels, ranging from 69 kV to 765 kV in transmission systems.


A branch is either a transmission line or a transformer. A transmission line can be thought of as a circuit that carries current and power between two points. The main function of a transformer is to connect buses at different voltage levels.  In both cases, it is described by a lumped circuit model with parameters $R$ (resistance) and $X$ (reactance). Together, they form the impedance  $Z=R+jX$, with the unit of Ohms. It is often easier to work with the inverse of $Z$, called the admittance, defined as $Y=\frac{1}{Z}$, with a unit of Siemens. For longer lines, the charging susceptances caused by the capacitance between the line and the ground are also included in the model. 

We also include HVDC (high voltage, direct current) lines in our models. They are used to transmit power across long distances where AC transmission is inconvenient or expensive. Instead of circuit elements, they are modeled as transport links that can carry power up to some capacity. 

At the organizational level, the continental US grid comprises three largely disconnected systems: the Eastern Interconnection, the Western Interconnection (WECC), and Texas. All the systems operate nominally at 60 Hz, but they are not synchronized. In each interconnection, there are a number of system operators and balancing authorities to ensure the balance of supply and demand in defined geographic areas. Next, we describe this balancing process in more detail. 
\subsection{Optimal Power Flow}\label{sec:opf-background}
System operators balance demand and supply by solving the optimal power flow (OPF) problem. More concretely, OPF finds the least cost solution subject to generation and demand balance and other network constraints.  In this section, we will first describe the fully nonlinear problem, called the AC-OPF problem, then we will describe some of its variants. 

\subsubsection{AC-OPF}
Consider two buses $i$ and $k$, connected by a transmission line with impedance $Z_{ik}$, or equivalently, admittance $Y_{ik}=\frac{1}{Z_{ik}}$. Let $V_i=|V_i| e^{j\theta_i}$ and $V_k=|V_k| e^{j\theta_k}$ be their voltages, respectively. The current from $i$ to $k$ is given by Ohm's law: $I_{ik}=Y_{ik} (V_i-V_k)$. The complex power from bus $i$ to bus $k$ is defined as 
$$ S_{ik}=V_i I_{ik}^*,$$
where $^*$ denotes the complex conjugate. Separating $S_{ik}$ into real and imaginary parts, we have
$$ S_{ik}=P_{ik}+j Q_{ik},$$
where $P_{ik}$ is called the active power\footnote{And is also called the real power in older texts} and $Q_{ik}$ is called the reactive power. Active power has units of watts (W) or, more commonly in practice, kilowatts (kW) or megawatts (MW).   Reactive power has units of volt-ampere-reactive (VAr), or kVAr and MVAr. 

There are several ways to write the power flow equations. Here, we provide the polar coordinate formulation. Writing $Y_{ik}=g_{ik}+j b_{ik}$, where $g_{ik}$ is called the conductance and $b_{ik}$ is called susceptance, we have 
\begin{equation} \label{eqn:line_flow}
\begin{split}
P_{ik} &= |V_i|^2 g_{ik} - |V_i||V_k|
          \left\{g_{ik}\cos(\theta_i-\theta_k)
          + b_{ik}\sin(\theta_i-\theta_k)\right\} \\
Q_{ik} &= -|V_i|^2 b_{ik} - |V_i||V_k|
           \left\{g_{ik}\sin(\theta_i-\theta_k)
           - b_{ik}\cos(\theta_i-\theta_k)\right\}
\end{split}. 
\end{equation}

The active power injection at bus $i$ is the sum of power flowing from bus $i$ to its neighbors, defined as $P_i=\sum_{k \sim i} P_{ik}$, where $k \sim i$ denote bus $k$ is connected to bus $i$. The reactive power injection $Q_i$ is defined in a similar way. Summing \eqref{eqn:line_flow}, we have
\begin{equation} \label{eqn:power_flow}
\begin{split}
P_i &= \sum_{k \sim i} |V_i|^2 g_{ik} - |V_i||V_k|
        \left\{g_{ik}\cos(\theta_i-\theta_k)
        + b_{ik}\sin(\theta_i-\theta_k)\right\} \\
Q_i &= \sum_{k \sim i} -|V_i|^2 b_{ik} - |V_i||V_k|
        \left\{g_{ik}\sin(\theta_i-\theta_k)
        - b_{ik}\cos(\theta_i-\theta_k)\right\}
\end{split},
\end{equation}
and these are called the power flow equations in polar coordinates. By convention, a positive $P_i$ (or $Q_i$) means that the bus is injecting power into the system.

We think of each device in the system as either a generator or a load.\footnote{This categorization is flexible, for example, a battery storage could be a generator or a load depending on whether it is charging or discharging.} We allow a bus to have both generation and load. For simplicity, we assume that a bus has at most one generator and at most one load. This assumption can be relaxed at the cost of slightly more cumbersome notations.

We assign a cost of generation at bus $i$, written as $C_i (P_{g,i})$, where $P_{g,i}$ is the active power generated and $C_i$ has units of \$. The load is typically modeled as given and has both active and reactive components, written as $P_{d,i}$ and $Q_{d,i}$. The AC-OPF problem is 
\begin{subequations} \label{eqn:ACOPF}
    \begin{align}
        \min\; &\sum_{i \in \mathcal{G}} C_i (P_{g,i})  \\
        \mbox{s.t. } 
        & P_i = \sum_{k \sim i} |V_i|^2 g_{ik} - |V_i||V_k| 
        \left\{g_{ik}\cos(\theta_i-\theta_k)
        + b_{ik}\sin(\theta_i-\theta_k)\right\} \label{eqn:P} \\
& Q_i = \sum_{k \sim i} -|V_i|^2 b_{ik} - |V_i||V_k|
        \left\{g_{ik}\sin(\theta_i-\theta_k)
        - b_{ik}\cos(\theta_i-\theta_k)\right\} \label{eqn:Q}\\
        & P_i=P_{g_i}-P_{d_i} \label{eqn:P_balance}\\
        & Q_i=Q_{g_i}-Q_{d_i} \label{eqn:Q_balance} \\
        & P_{g,i}^{\min} \leq P_{g_i} \leq P_{g,i}^{\max} \label{eqn:P_gen} \\
        & Q_{g,i}^{\min} \leq Q_{g_i} \leq Q_{g,i}^{\max} \label{eqn:Q_gen}\\
        & V_i^{\min} \leq |V_i| \leq V_i^{\max} \label{eqn:V} \\
        & |S_{ik}| \leq S_{ik}^{\text{rated}} \label{eqn:S}\\
        & |\theta_i - \theta_k| \leq \theta_{ik}^{\max}, \label{eqn:theta} 
    \end{align}
\end{subequations}
where \eqref{eqn:P} and \eqref{eqn:Q} are the power flow equations, \eqref{eqn:P_balance} and \eqref{eqn:Q_balance} are the power balance equations (supply=demand), \eqref{eqn:P_gen} and \eqref{eqn:Q_gen} are the generator limits, \eqref{eqn:V} is the voltage limit, \eqref{eqn:S} is the line thermal limits, and \eqref{eqn:theta} is the angle limit coming from stability constraints. 

Overall, the AC-OPF problem is nonconvex and NP-hard in the worst case~\cite{Low_draft}. The study of this problem is a central subject in power systems and the interested reader can consult several standard textbooks and papers~\cite{Low_draft,bergen2009power,zhang2012geometry,low2014convex1,low2014convex2}. In practice, the AC-OPF problem is generally solvable if it is feasible. More precisely, several solvers (e.g., IPOPT~\cite{wachter2006}) can solve networks of up to $\sim10,000$ buses in minutes, provided the problem has a solution. If the demand comes from actual measurements or historical data, then feasibility of AC-OPF is not typically a critical issue. 

However, for generated networks, the feasibility of AC-OPF can be a critical blocker. We are essentially asking for a set of data that satisfies the set of nonlinear equalities and inequalities under the constraint of \eqref{eqn:ACOPF}. A generic set of parameters is typically not feasible, and it is not obvious how to make them feasible while ensuring that they are still realistic. An important contribution of this paper is to provide an interpretable and implementable method to accomplish this. 

We end this subsection by explaining a key simplification of the AC-OPF problem, called DC-OPF.\footnote{This terminology is unfortunate since the equations have nothing to do with direct current. It comes from the historical fact that DC analyzers were used to approximate AC power flow, and the name has stuck.} The DC power flow equations are a linearization of the AC power flow equations, where we set all voltage magnitudes to be $1$ per unit, all conductances ($g_{ik}$) to 0, and assume angles differences are small so $\sin(\theta_i-\theta_k) \approx \theta_i-\theta_k$. With these assumptions, reactive powers are constant, and \eqref{eqn:ACOPF} simplifies to:
\begin{subequations} \label{eqn:DCOPF}
    \begin{align}
        \min\; &\sum_{i \in \mathcal{G}} C_i (P_{g,i})  \\
        \mbox{s.t. } 
        & P_i = \sum_{k \sim i} -b_{ik}(\theta_i-\theta_k) \\
        & P_i=P_{g_i}-P_{d_i} \\
        & P_{g,i}^{\min} \leq P_{g_i} \leq P_{g,i}^{\max} \\
        & |\theta_i - \theta_k| \leq \theta_{ik}^{\max}. 
    \end{align}
\end{subequations}
This problem is a linear program and thus can be easily solved. Because losses are ignored, the cost of DC-OPF is typically slightly lower than the cost of AC-OPF (by around~5\%)~\cite{molzahn2019}.

\subsection{OpenStreetMap as a Data Source}\label{sec:osm}

OpenStreetMap (OSM) is a collaborative geographic database with over
10~million registered users~\citep{osm_stats} and tens of millions of elements
tagged with the \texttt{power=*} key
worldwide~\citep{osm_taginfo}.  The power tagging schema provides a
structured vocabulary for mapping electrical infrastructure:
\texttt{power=line} (overhead transmission lines, with voltage,
conductor count, and circuit attributes), \texttt{power=cable}
(underground/undersea cables), \texttt{power=substation},
\texttt{power=generator}/\texttt{plant} (with fuel type and capacity),
\texttt{power=converter} (HVDC terminals), and
\texttt{power=transformer}.

For the United States, OSM coverage of high voltage transmission
(345~kV and above) is substantial.  Coverage diminishes at lower
voltages: subtransmission (69--161~kV) is unevenly mapped, and
distribution ($<$69~kV) is largely absent -- acceptable for
transmission-level studies.  The OSM power data was validated in~\cite{arderne2020} and the authors found that it is sufficient to reconstruct high-voltage networks in most developed countries.

OSM data have four fundamental limitations for power systems modeling:
(1)~the lack of electrical parameters (e.g., line impedances and thermal ratings), (2)~missing parallel circuits (a multi-circuit corridor appears as a single geographic feature), (3)~incomplete voltage tags (e.g., about 15--30\% of US lines do not have tags), and (4)~no demand or cost data.  Our solutions in~\cref{sec:topology}, \cref{sec:parameters}, and \cref{sec:demand} address these challenges and close the data gap. 


\paragraph{Legal context.}
It is important to note that CEII protections cover proprietary network models -- internal bus
numbering, measured impedance values, control settings, and
market-sensitive information.  They \textbf{do not prohibit} mapping physical infrastructure that is plainly visible from public roads, satellite imagery, or other sources. Transmission towers, substations, and power plants are prominent features of the built environment.  OpenStreetMap contributors map them the same way they map roads and buildings.  In our work, we derive grid models entirely from publicly observable geographic data and government statistics; we do not access or reverse-engineer any CEII-protected data. For example, the electrical parameters we assign (see \cref{sec:parameters}) are engineering estimates from textbook lookup tables, not measured values from protected models.

\section{System}\label{sec:system}

\subsection{Pipeline Architecture}\label{sec:architecture}

Our pipeline transforms raw OpenStreetMap data into solver-ready OPF
models through five sequential stages (\cref{tab:pipeline-stages}). After the models are constructed, we solve OPF problems using PowerModels.jl~\citep{coffrin2018}, a Julia framework for power systems optimization that interfaces with the
Ipopt~\citep{wachter2006} interior-point solver.


\begin{table}[h!]
\centering\small
\caption{Pipeline stages overview.}
\label{tab:pipeline-stages}
\begin{tabular}{@{}clll@{}}
\toprule
\textbf{Step} & \textbf{Stage} & \textbf{Primary Input} & \textbf{Output} \\
\midrule
1 & Data Extraction    & OSM (Overpass API)         & GeoJSON per state \\
2 & Topology Reconstruction          & GeoJSON                    & Bus-branch network \\
3 & Parameter Estimation     & Network + LUTs + EIA       & Parameterized model \\
4 & Demand Allocation  & EIA-930 + Census + HIFLD   & Model with loads \\
5 & OPF Solve          & Complete model             & Solved power flow \\
\bottomrule
\end{tabular}
\end{table}

\paragraph{Data sources.} Here, we describe the input data sources to our pipeline. They are all publicly and freely available:
\begin{itemize}[leftmargin=*,noitemsep]
\item \textbf{OpenStreetMap (OSM)}~\citep{osm2026}: infrastructure geography
  (substations, transmission lines, generators).
\item \textbf{U.S.\ Energy Information Administration (EIA)
  Form~860}~\citep{eia860_2024}: generator inventory -- fuel
  type, nameplate capacity, and operating status for every US power
  plant operating at $\geq 1 \ \text{MW}$, used to validate and supplement OSM generator data.
\item \textbf{EIA-923}~\citep{eia923_2024}: generator heat rates and
  fuel costs, used to build quadratic cost curves.
\item \textbf{EIA-930}~\citep{eia930_2026}: hourly demand of each balancing
  authority (BA) in the US. This is the primary input for the demand allocation stage.
\item \textbf{EIA Electric Power Annual}~\citep{eia_epa2024}:
  circuit-miles of transmission line by voltage class, used to calibrate
  topology and capacity scaling factors.
\item \textbf{EIA Natural Gas Prices}: Henry Hub spot price (\$/MMBtu),
  used to override default natural gas fuel prices in generator cost
  curves so that dispatch reflects current market conditions.
\item \textbf{US Census Bureau}~\citep{census2024}: census-tract
  population counts, used to distribute BA-level demand to individual buses via scaling proportionally to population~\cite{taylor1975demand,hyndman2009density}. 
\item \textbf{Electric planning Homeland Infrastructure Foundation-Level
Data (HIFLD)}~\citep{hifld_ba}: Balancing authority boundary polygons, used to
  assign buses to BAs via spatial containment.
\end{itemize}

\paragraph{Output format.}
The final model is exported as a PowerModels-compatible JSON file
(MATPOWER structure) for the OPF solver, with all values in per-unit on
a 100~MVA base (see Appendix~\ref{app:perunit} for more details).

\subsection{Data Extraction (Step 1)}\label{sec:extraction}

The extraction stage queries a local Overpass~API instance for
power infrastructure features matching a defined set of tag values within a given US state, and converts them into a
normalized GeoJSON feature collection.  For networks spanning multiple states,
per-state extracts are merged with cross-border de-duplication.

\subsubsection{OSM Power Tagging Schema}

OpenStreetMap represents geographic features using three primitive types:
\emph{nodes} (points), \emph{ways} (ordered sequences of nodes forming
lines or polygons), and \emph{relations} (groupings of nodes and ways).
Each element carries key-value \emph{tags} that describe its properties.
The power infrastructure tagging schema~\citep{osm2026} defines a
structured vocabulary under the \texttt{power=*} key.
\Cref{tab:osm-tags} lists the feature types used by the pipeline.

\begin{table}[h!]
\centering\small
\caption{OSM power feature types used in the pipeline.}
\label{tab:osm-tags}
\begin{tabular}{@{}llp{5.2cm}@{}}
\toprule
\textbf{OSM Tag} & \textbf{Geometry} & \textbf{Key Attributes} \\
\midrule
\texttt{power=line}        & Way   & \texttt{voltage}, \texttt{cables},
                                       \texttt{circuits}, \texttt{frequency} \\
\texttt{power=cable}       & Way   & \texttt{voltage}, \texttt{cables},
                                       \texttt{location} \\
\texttt{power=substation}  & Node/Polygon & \texttt{voltage},
                                       \texttt{substation} (type) \\
\texttt{power=plant}       & Node/Polygon & \texttt{plant:output:electricity},
                                       \texttt{plant:source} \\
\texttt{power=converter}   & Node  & \texttt{voltage}, \texttt{converter (HVDC)} \\
\bottomrule
\end{tabular}
\end{table}

OSM also defines \texttt{power=generator} (individual turbine/panel
nodes within a plant) and \texttt{power=} \texttt{transformer} tags.  The pipeline
downloads both but does not use them directly. Instead, generators are modeled at
the aggregated \texttt{plant} level to avoid duplicate entries at a bus, and transformers are inferred from voltage transitions
at substations rather than from the sparsely tagged OSM elements
(in our US extract, fewer than 20\% of inferred transformer
locations carry an explicit \texttt{power=transformer} tag).
Section \ref{sec:topology} provides more details on the inference approach.

The \texttt{voltage} tag is the most important attribute for our grid modeling process.  It is specified in volts (e.g., \texttt{345000} for 345~kV)
and may be semicolon-delimited for multi-voltage features (e.g.,
\texttt{345000;138000}).  The \texttt{cables} tag counts physical
conductors, while \texttt{circuits} counts independent electrical
circuits. Since transmission lines are balanced three-phase, a single three-phase circuit uses three cables and a double-circuit tower carries six.  The
\texttt{frequency} tag distinguishes AC transmission lines from high voltage DC lines (see Section~\ref{sec:hvdc} for more details).


Generator capacity and fuel type are primarily sourced from
EIA-860 (Section~\ref{sec:parameters}) and the OSM \texttt{plant} tags provide a
geographic starting point and cross-validation reference.

\subsubsection{Local Overpass API}

The Overpass API is a read-only query interface for OSM data that
supports spatial and attribute-based filtering.  Public Overpass
instances enforce rate limits that make bulk extraction of US-scale data
impractical.  We therefore deploy a local instance using Docker, loaded
with the official US extract (${\sim}10$~GB compressed).

Our query uses area-based clipping for precise state boundary alignment
rather than bounding boxes, which would include features from
neighboring states. 
The downloader produces a GeoJSON preserving all
original OSM tags as feature properties.

\subsubsection{Data Quality Assessment}\label{sec:data-quality}

Before proceeding to topology reconstruction, we assess what OSM
captures and what it misses by comparing OSM-derived statistics against
the EIA Electric Power Annual~\citep{eia_epa2024}, which reports
circuit-miles of transmission line by voltage class.

\paragraph{Route-miles vs.\ circuit-miles.}
OSM maps transmission line \emph{routes} -- the physical path a corridor
follows.  The EIA reports \emph{circuit-miles}, counting each parallel
circuit separately (a 100-mile corridor with 4~circuits counts as
400~circuit-miles).  Since OSM typically represents each corridor as a
single way regardless of circuit count, comparing the two provides some information on the number of parallel circuits, as shown by the ratio between the OSM count and the EIA count in Table~\ref{tab:osm-eia}.

\begin{table}[h!]
\centering\small
\caption{OSM route-miles vs.\ EIA circuit-miles by voltage class
(48~continental US states).}
\label{tab:osm-eia}
\begin{tabular}{@{}rrl@{}}
\toprule
\textbf{Voltage (kV)} & \textbf{OSM\,/\,EIA} & \textbf{Interpretation} \\
\midrule
  69  & 1.5$\times$ & Way fragmentation \\
 115  & 2.9$\times$ & Significant over-count \\
 138  & 2.1$\times$ & Over-count \\
 161  & 4.1$\times$ & Likely tagging inconsistencies \\
 230  & 1.5$\times$ & Moderate over-count \\
 345  & 1.8$\times$ & Over-count \\
 500  & 2.3$\times$ & Over-count \\
 765  & 0.5$\times$ & \textbf{Under-mapped} \\
\bottomrule
\end{tabular}
\end{table}

At most voltage levels, OSM-derived branch-miles exceed EIA
circuit-miles.  Two artifacts drive the inflation: (1)~multi-voltage
corridor splitting  --  a way tagged \texttt{345000;138000} produces two
branch records, each carrying the full corridor length  --  and
(2)~cross-state double-counting, since each state is processed
independently and border corridors appear in both models.
These ratios are therefore upper bounds on actual route-miles;
after accounting for double counting, the OSM coverage is broadly
consistent with EIA totals at 230--345~kV.
The exception is 765~kV, where OSM captures only about
half the expected mileage, likely because some high voltage
lines in remote areas have not yet been mapped.

\paragraph{Missing voltage tags.}
Across the 48 continental states, approximately 15--30\% of
\texttt{power=line} ways lack a \texttt{voltage} tag.  This fraction is
higher for subtransmission and lower for EHV lines ($\ge$345~kV), with the latter 
tend to be mapped by more experienced contributors.  The voltage
inference algorithm described in Section~\ref{sec:topology} addresses this gap.

\paragraph{Missing parallel circuits.}
OSM provides no reliable way to determine how many parallel circuits
share a physical corridor.  The \texttt{circuits} tag is often not present and the \texttt{cables} tag counts physical conductors, not circuits.  This gap is the primary motivation for the
topology and capacity scaling factors described in Section~\ref{sec:topology-factors}.

\subsection{Topology Reconstruction (Step 2)}\label{sec:topology}

Topology reconstruction transforms the raw GeoJSON feature collection
into a structural bus-branch model -- buses with voltage levels,
branches with geographic routes, generators assigned to buses, and HVDC
links identified.  Electrical parameters (impedance, thermal ratings,
cost curves) are deferred to the parameter estimation stage
(\cref{sec:parameters}).

The reconstruction proceeds through a sequence of sub-steps, summarized
in \cref{tab:topo-steps}.

\begin{table}[h!]
\centering\small
\caption{Topology reconstruction sub-steps.}
\label{tab:topo-steps}
\begin{tabular}{@{}cl@{}}
\toprule
\textbf{\#} & \textbf{Task} \\
\midrule
 1 & Load and parse OSM GeoJSON (\cref{sec:load-parse}) \\
 2 & Infer missing voltages (\cref{sec:voltage-inference}) \\
 3 & Filter by voltage $\ge 69$~kV (\cref{sec:voltage-filter}) \\
 4 & Parse circuit counts (\cref{sec:circuit-parsing}) \\
 5 & Build facility footprints (\cref{sec:facility-footprints}) \\
 6 & Build spatial endpoint index (\cref{sec:endpoint-index}) \\
 7 & Merge fragmented line segments (\cref{sec:line-merging}) \\
 8 & Classify circuits (\cref{sec:circuit-class}) \\
 9 & Create buses (\cref{sec:bus-creation}) \\
10 & Detect transformers (\cref{sec:bus-creation}) \\
11 & Detect HVDC links (\cref{sec:hvdc}) \\
12 & Assign generators to nearest bus (\cref{sec:gen-assign}) \\
13 & Validate and extract largest component (\cref{sec:validation}) \\
\bottomrule
\end{tabular}
\end{table}

A typical single-state model contains 100--4{,}000 buses,
4--350 generators, and 200--6{,}500 branches.  For the remainder of
this section, we use Virginia as a running example to illustrate each
sub-step with concrete numbers.

\subsubsection{Load and Parse}\label{sec:load-parse}

The GeoJSON feature collection produced by the extraction stage
(\cref{sec:extraction}) is loaded and split by \texttt{power} type into
substations, lines (including cables), and generators. Geometric objects that are not lines can sometimes appear in the lines layer (polygons and
points) and are filtered out, voltages are parsed from semicolon-delimited
strings into numeric lists, and generator capacities are normalized to
MW.  Additionally, HVDC status flags are set on each line using
tag-based signals (\cref{sec:hvdc}), so that subsequent circuit parsing
can separate AC and DC circuits from the outset.

\subsubsection{Voltage Inference}\label{sec:voltage-inference}

Since voltage determines which lines can be
electrically connected and drives all parameter estimation
(\cref{sec:parameters}), filling the gaps identified in
\cref{sec:data-quality} is an essential early step.

The algorithm performs iterative neighbor consensus over up to
10~iterations.  In each iteration, every untagged line segment is
examined against its topological neighbors -- other segments that share a
co-located endpoint.  Voltage is assigned by the first matching rule:

\begin{enumerate}[leftmargin=*]
\item \textbf{Unanimous agreement.}  All neighbors at an endpoint share
  the same voltage.
\item \textbf{Supermajority.}  At least three neighbors exist and
  $\geq 2/3$ agree on a voltage.
\end{enumerate}

\begin{figure}[h!]
\centering
\includegraphics[width=0.9\columnwidth]{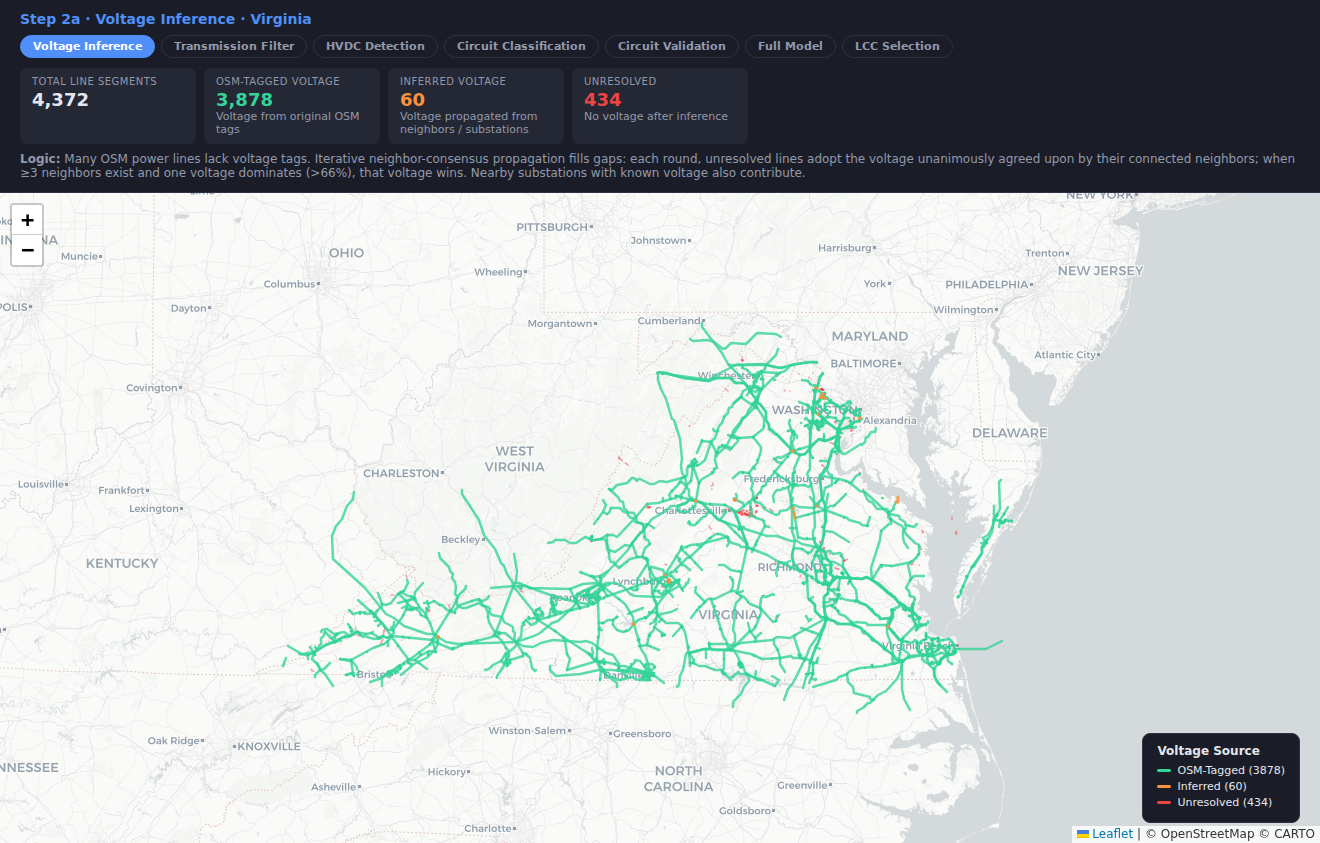}
\caption{Voltage inference for Virginia: 3{,}878 OSM-tagged lines
  (green), 60 inferred by neighbor consensus (orange), and 434
  unresolved (red).}
\label{fig:topo-va-voltage}
\end{figure}

\noindent
When an endpoint falls within a substation whose own \texttt{voltage}
tag is set, that voltage is added to the candidate pool alongside
neighbor votes, improving consensus at junctions where few line
segments are tagged.

\noindent
Convergence typically occurs within 3--5~iterations.  For Virginia,
the algorithm starts with 4{,}372 line segments, of which 3{,}878
already carry voltage tags; iterative inference resolves 60 more,
leaving 434 ($\sim$10\%) unresolved (see \cref{fig:topo-va-voltage}).

\subsubsection{Voltage Filter}\label{sec:voltage-filter}

\begin{figure}[h!]
\centering
\includegraphics[width=0.9\columnwidth]{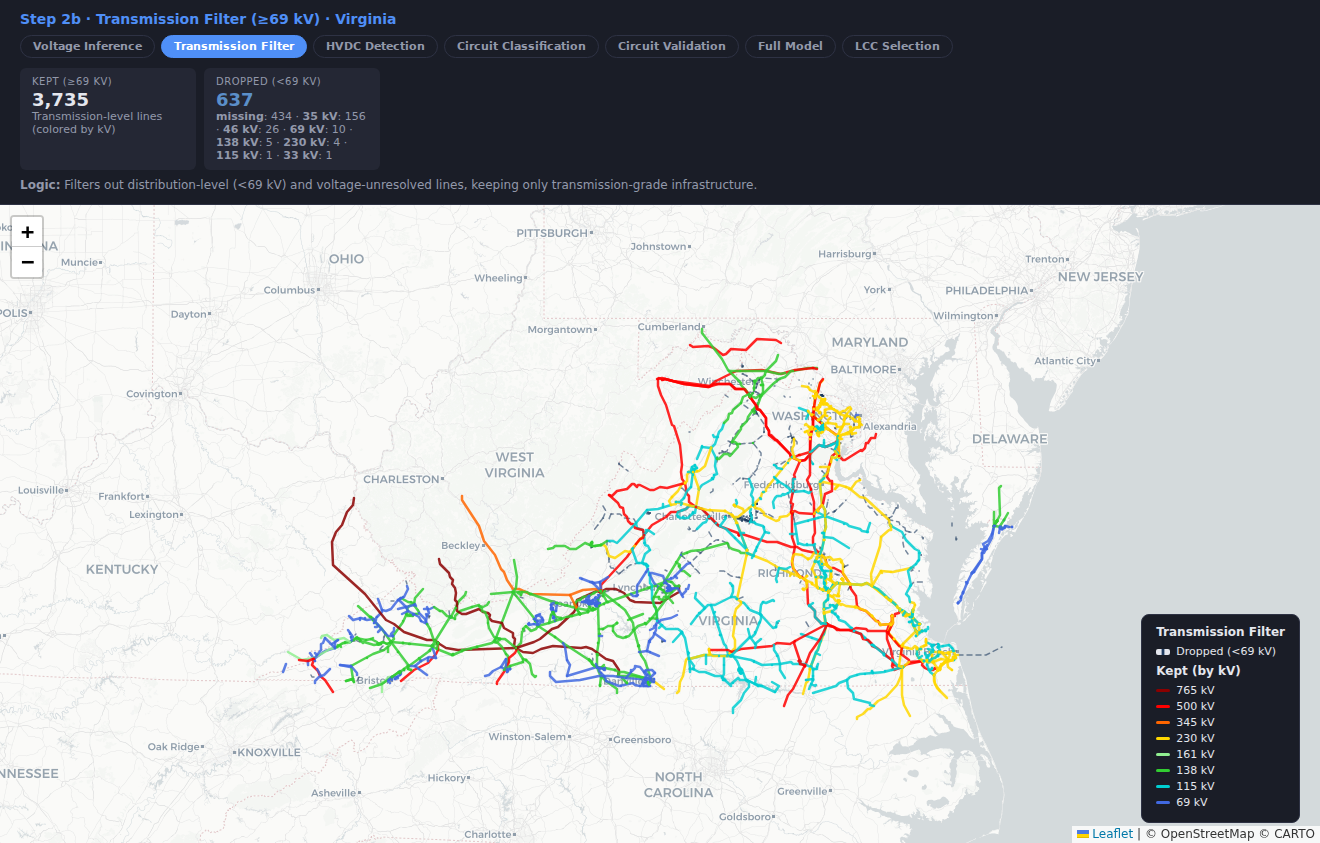}
\caption{Transmission filter ($\ge$69~kV) for Virginia: 3{,}735
  segments retained, colored by voltage class (69~kV blue through
  765~kV brown).}
\label{fig:topo-va-filter}
\end{figure}

Following NERC's Bulk Electric System definition, all features below 69~kV are removed, separating bulk transmission from sub-transmission and distribution.  Lines whose voltage could not be
resolved by inference are also dropped.  For Virginia,
this step drops 637 of 4{,}372 segments, retaining 3{,}735
transmission-grade line segments (\cref{fig:topo-va-filter}).

\subsubsection{Circuit Count Parsing}\label{sec:circuit-parsing}

A single OSM way may carry multiple electrical circuits -- for example, a
double-circuit tower supports two independent three-phase circuits on
the same structure, and a multi-voltage corridor may carry circuits at
345~kV and 138~kV side by side.  Correctly determining the number of
circuits per way is essential in the bus-branch model.

OSM provides three tags that carry circuit-count information, but they
are inconsistently applied and sometimes contradictory:
\begin{itemize}[leftmargin=*]
\item \texttt{circuits}: the most direct indicator, but present on only
  $\sim$19\% of ways in our US extract.
\item \texttt{cables}: the number of physical conductors, present on
  $\sim$88\% of ways in our US extract.  For a three-phase AC circuit, three cables form
  one circuit, so \texttt{cables=6} implies two circuits.
\item \texttt{voltage}: a semicolon-delimited list (e.g.,
  \texttt{345000;138000}) whose length indicates distinct circuits at
  different voltage levels.
\end{itemize}

The pipeline reconciles these signals in two passes.  A first pass
applies a simple priority rule: explicit \texttt{circuits} tag
$\rightarrow$ \texttt{cables} $\div$ 3 $\rightarrow$ default of~1.
A second refinement pass resolves disagreements between the declared
circuit count~$C$, the voltage-list length~$V$, and the cable count
using a configurable mode.  The default mode, \emph{trust\_voltage},
lets the voltage count override when $V \neq C$: a way tagged
\texttt{voltage=345000;138000} produces two circuit records -- one per
voltage level -- regardless of what the \texttt{circuits} or
\texttt{cables} tags say.  When $V > C$, the extra circuits are
assigned to the highest voltages first; when $V < C$, additional
circuits inherit the single declared voltage.

Multi-voltage splitting is critical for parameter estimation: a 345~kV
circuit and a 138~kV circuit on the same corridor have different
impedance and thermal ratings (\cref{sec:parameters}), and must be
modeled as separate branches.  HVDC circuits receive a distinct key
suffix so that AC and DC circuits are never merged in subsequent
processing steps.

\subsubsection{Facility Footprints}\label{sec:facility-footprints}

Substation polygons and generator plant polygons are collected into a
spatial index of facility areas.  Polygons are expanded by a small
buffer ($0.0006^{\circ}$, $\approx$66~m) to account for endpoints that
fall just outside the digitized boundary.  Substations mapped as points
(rather than area polygons) are assigned a distance-based buffer
($\sim$100~m) at query time.  Line endpoints are later tested
against these footprints to determine which facility they belong to.

\subsubsection{Endpoint Index}\label{sec:endpoint-index}

All line endpoints are discretized onto an integer grid at
$10^{-6}$ degree resolution ($\approx$11~cm) and stored in a hash-based
spatial index.  Each endpoint is annotated with the facility it falls
within (if any), enabling the subsequent merging and classification
steps to operate in near-linear time.

\subsubsection{Line Merging}\label{sec:line-merging}

A single physical transmission line between two substations is typically
fragmented across many OSM ways (one per span between towers, or split
at mapper boundaries).  A 100-mile line may comprise 20--50 ways.  These
need to be stitched together to recover complete circuits.

All way endpoints are snapped to the same $10^{-6}$ degree grid used
by the endpoint index (\cref{sec:endpoint-index}).  A union-find
data structure then merges ways whose snapped endpoints coincide and
whose voltages are compatible (identical, or one inherits the other's).
After merging, each disjoint set represents a continuous circuit between
two facilities.  Merging typically reduces the feature count by
60--80\%. For example, Virginia's 3{,}735 line segments merge into 1{,}644
distinct circuits.

\subsubsection{Circuit Classification}\label{sec:circuit-class}

Each merged circuit is classified by endpoint connectivity:

\begin{itemize}[leftmargin=*]
\item \textbf{Inter-facility}: endpoints at two different substations or
  generation sites -- the only type retained for the OPF model.
\item \textbf{Loop}: both endpoints at the same facility (internal
  busbar wiring).
\item \textbf{Self-loop}: a degenerate merge artifact in which the same
  line section index appears more than once in a merged circuit group,
  detected and removed before geometry reconstruction.
\item \textbf{Single-facility}: one endpoint at a facility, the other
  dangling.
\item \textbf{Isolated}: neither endpoint near any facility.
\item \textbf{Tap}: a T-junction spur off a main line at a tower.
\end{itemize}

\begin{figure}[h!]
\centering
\includegraphics[width=\columnwidth]{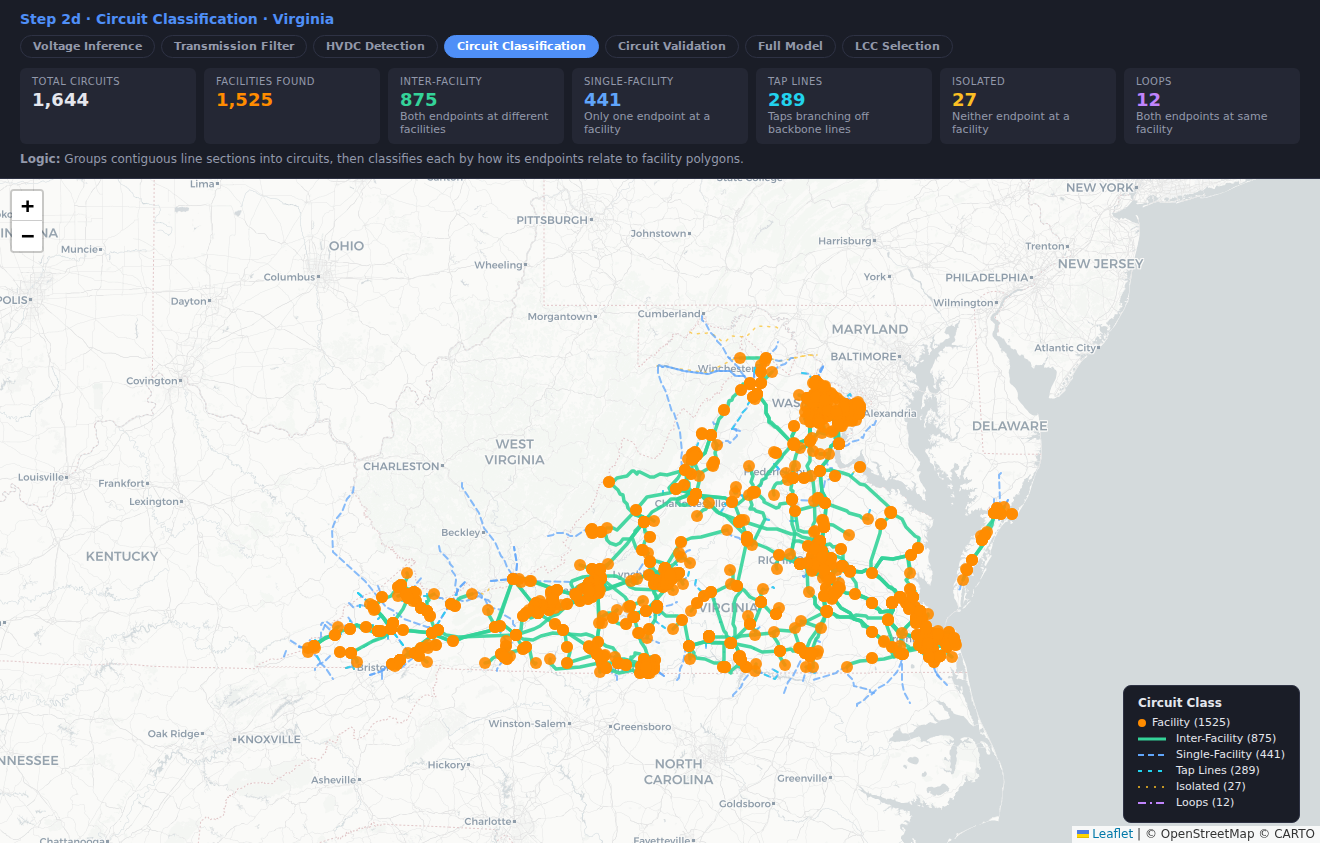}
\caption{Circuit classification for Virginia: 875 inter-facility
  circuits (green solid), 441 single-facility (dashed), 289 taps, 27 isolated, and 12 loops.  Only
  inter-facility circuits become branches in the OPF model.}
\label{fig:topo-va-classify}
\end{figure}

\noindent
Only inter-facility circuits become branches in the bus-branch model;
removing taps biases the model toward meshed transmission behavior and
may under-represent last-mile congestion into load pockets.
For Virginia, 875 of 1{,}644 circuits are inter-facility; the remainder
are loops~(12), single-facility~(441), taps~(289), or
isolated~(27) -- mapping artifacts that carry no inter-substation power
flow (\cref{fig:topo-va-classify}).

\subsubsection{Bus Creation and Transformer Detection}\label{sec:bus-creation}

\paragraph{Spatial clustering.}
Line endpoints that fall within a substation footprint are assigned to
that substation.  Endpoints outside any mapped substation are clustered
using a union-find approach with a $0.0005^{\circ}$ threshold
($\approx$50~m) to form ad-hoc bus locations.

\paragraph{Multi-voltage splitting.}
Most substations operate at multiple voltage levels (e.g.,
345/138/69~kV).  Because each voltage level must be a separate bus in
the OPF model, the pipeline splits a spatial cluster into distinct buses
when incident circuits span voltages that differ by more than 20\%.
This conservative threshold avoids splitting voltage classes that are
nominally distinct but close.

\paragraph{Transformer inference.}
At any substation where buses at different voltage levels coexist, a
transformer branch is inferred between the corresponding bus pair.
A stricter dual threshold (${>}\,10$~kV absolute difference \emph{and}
${>}\,1.2{\times}$ ratio) is used here to suppress false positives from
minor tagging discrepancies (e.g., 230/220~kV).  In our US OSM extract,
fewer than 20\% of inferred transformer locations carry an explicit
\texttt{power=transformer} tag, making inference the primary detection
method.  A final catch-all pass converts any remaining branch whose
endpoint buses differ by more than 10\% in voltage into a transformer,
using the loosest threshold to avoid leaving any clear voltage mismatch
as an AC line.  Transformer electrical parameters
are assigned in \cref{sec:parameters}.

\subsubsection{Generator Assignment}\label{sec:gen-assign}

Generators (from OSM \texttt{plant} features, supplemented by EIA-860)
are assigned to their nearest bus via a spatial join with a maximum
distance of ${\approx}\,1$~km.  When an EIA-860 match is available
(by name, fuel type, and proximity), the generator's capacity is updated
with the official EIA value.  Default initial dispatch is set to 50\% of
$P_{\max}$ as a placeholder; this value is overwritten by the merit-order dispatch
in \cref{sec:dispatch} before the OPF solver runs.

\subsubsection{HVDC Detection}\label{sec:hvdc}

HVDC lines are electrically distinct from AC branches and connect to the AC network through converter stations.  However, OSM has no dedicated HVDC tag; these lines are tagged as ordinary \texttt{power=line} or \texttt{power=cable}
features.  The pipeline therefore infers HVDC status using OR logic
across multiple independent signals -- any one is sufficient:
\begin{enumerate}[leftmargin=*]
\item \texttt{frequency=0} or \texttt{frequency=dc}.
\item Voltage tag contains a $\pm$ prefix (bipolar HVDC convention).
\item \texttt{line:type} or \texttt{cable:type} set to \texttt{dc}.
\item Cable count consistent with DC (1 or 2 conductors at
  ${>}\,100$~kV, with no AC frequency tag).
\item Feature name matches a curated list of known US HVDC projects
  (e.g., Pacific Intertie, Cross-Sound Cable).
\end{enumerate}

\noindent
A separate post-processing pass checks whether any remaining
unclassified line has both endpoints within 500~m of a
\texttt{power=converter} node, catching cases where none of the tag-based
signals are present.

Detected HVDC lines are exported as \texttt{dcline} entries in the
PowerModels format, modeled as controllable point-to-point links with
active-power limits derived from voltage class.  Loss model parameters
and reactive-power limits are assigned in \cref{sec:parameters}.

\subsubsection{Validation and Final Assembly}\label{sec:validation}

The assembled network undergoes several validation and clean-up steps
before export:

\begin{enumerate}[leftmargin=*]
\item \textbf{Voltage-level bridging.}  At multi-voltage substations
  where separate voltage-level buses are disconnected (no inferred
  transformer yet links them), a transformer bridge is created to
  restore connectivity.  EHV substations ($\ge$345~kV)
  receive two parallel units for N-1 redundancy.  Electrical
  parameters for these bridges are assigned alongside all other
  transformers in \cref{sec:parameters}.
\item \textbf{Isolated bus removal.}  Buses with no connected branches
  are removed.
\item \textbf{Disconnected component removal.}  Disjoint connected components with no generators are removed since loads cannot be met. 
\item \textbf{Largest connected component.}  OPF requires a single
  connected AC network; the largest component (typically 90--99\% of
  buses) is retained.
\item \textbf{Slack bus assignment.}  The bus hosting the generator with
  the largest $P_{\max}$ is designated as the reference (slack) bus.
\end{enumerate}

\begin{figure}[h!]
\centering
\includegraphics[width=\columnwidth]{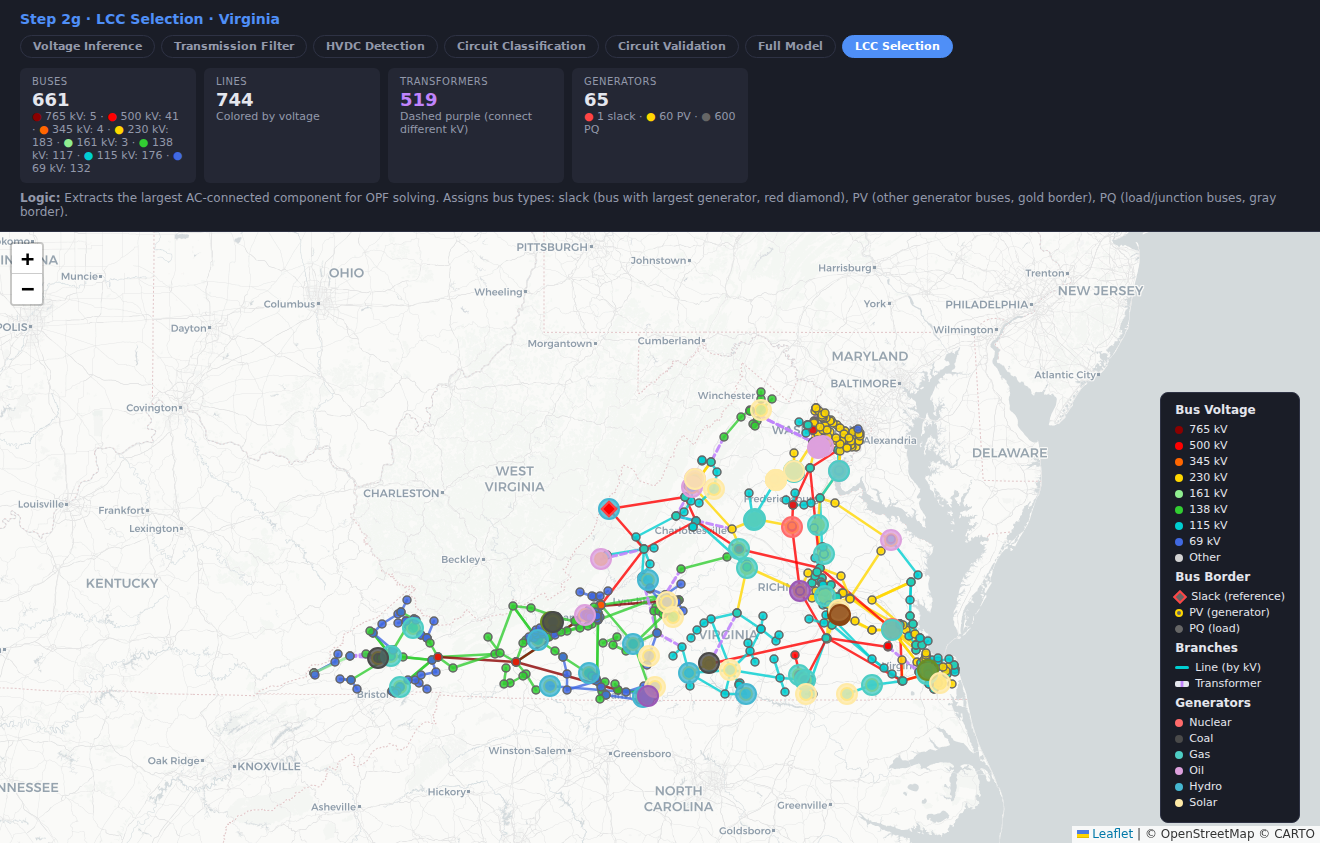}
\caption{Final bus-branch model for Virginia (largest connected
  component): 661 buses colored by voltage class, 744 AC lines,
  519 inferred transformers (dashed purple), and 65 generators
  sized by capacity and colored by fuel type.}
\label{fig:topo-va-lcc}
\end{figure}

\noindent
For Virginia, the full network contains 703 buses across
13~connected components; the largest component retains 661 buses
(94\%), 744 AC lines, 519 inferred transformers, and 65~generators.
The reduction from 875 inter-facility circuits to 744 AC lines
reflects isolated-bus and orphan-component pruning,
and voltage-level bridging adjustments during validation (\cref{fig:topo-va-lcc}).

\subsection{Parameter Estimation (Step 3)}\label{sec:parameters}

The topology reconstruction stage (\cref{sec:topology}) produces a
bus-branch model with geographic coordinates, voltage levels, and
structural connectivity -- but no electrical parameters.  This section
assigns the quantitative values that every OPF formulation
requires: line impedance and thermal ratings
(\cref{sec:line-params}), transformer impedance
(\cref{sec:xfmr-params}), parallel-circuit scaling factors
(\cref{sec:topology-factors}), generator cost curves and operational
limits (\cref{sec:gen-costs}), and reactive-power limits
(\cref{sec:reactive}).

\subsubsection{Line Parameter Estimation}\label{sec:line-params}

OSM provides the geographic route and voltage of a transmission line but
nothing about its electrical characteristics.  Resistance~($R$),
reactance~($X$), shunt susceptance~($B$), and thermal rating~(MVA) are
therefore estimated from the line's voltage class and assumed conductor
type using voltage-indexed lookup tables (LUTs) derived from standard
power engineering references~\cite{glover2012power} and conductor
manufacturer catalogs.

Each voltage class is associated with a representative conductor
configuration reflecting typical US utility practice
(see Appendix \ref{app:lut}).  At extra-high voltages ($\ge$345~kV), conductors are
bundled (2--4 sub-conductors per phase) to reduce corona discharge and
increase current-carrying capacity.  The bundle configuration
significantly affects impedance: a quad-bundle 765~kV line has very low
resistance and high surge-impedance loading ($X/R \approx 35$), while a
single-conductor 69~kV line is resistance-dominant ($X/R \approx 8$).
All resistance values are at 75\textdegree C, the standard operating
temperature for thermal-limit studies.

Lines identified as underground cables via \texttt{power=cable} or
\texttt{location=underground} tags use a separate LUT with
characteristics reflecting XLPE
insulation: lower reactance (close phase spacing), much higher shunt
susceptance (insulation dielectric), and lower thermal ratings (limited
by insulation temperature rather than ambient cooling).

Raw impedance values in $\Omega$/km are converted to per-unit on a
100~MVA system base using
\begin{equation}\label{eq:zpu}
  Z_\text{pu} = \frac{Z_\Omega \times L_\text{km}}{V_\text{kV}^2 / S_\text{base}},
\end{equation}
where $L_\text{km}$ is the line length computed from the merged
circuit's GeoJSON geometry (geodesic distance).

LUT thermal ratings represent continuous (normal) ratings.  Since each
OPF snapshot represents a single hour, short-term ratings are more
appropriate than continuous limits.  Transmission owners typically
establish short-term emergency ratings 10--15\% above the continuous
value, reflecting allowable transient conductor heating over periods of
15~minutes to 4~hours.  The pipeline applies a configurable thermal
margin of $1.10\times$ to all branch MVA limits to approximate
short-term ratings.

\subsubsection{Transformer Parameters}\label{sec:xfmr-params}

Each inferred transformer (\cref{sec:bus-creation}) is assigned
impedance and rating from a lookup table indexed by its high-voltage /
low-voltage pair (appendix \ref{app:lut}).  The table covers 52 voltage-pair
combinations, with typical values of
$X_\text{pu} = 0.05$--$0.16$ and
$R_\text{pu} = 0.002$--$0.008$ on the transformer's own MVA base.

For auto-transformers -- common when the voltage ratio is less than 3:1
and both sides are $\ge$230~kV (e.g.\ 345/230~kV) -- the impedance is
reduced by a winding-sharing factor:
\begin{equation}\label{eq:auto}
  Z_\text{auto} = Z_\text{base} \times
    \left(1 - \frac{V_\text{LV}}{V_\text{HV}}\right),
  \quad\text{clamped to } [0.20,\;0.65].
\end{equation}
This reflects the physical reality that an auto-transformer's impedance
is proportional to the voltage \emph{difference} rather than the ratio,
since only part of the winding carries the full current.

\subsubsection{Topology and Capacity Factors}\label{sec:topology-factors}

The most significant calibration challenge is compensating for missing
parallel circuits.  As discussed in \cref{sec:circuit-parsing}, OSM
represents each transmission corridor as a single geographic route
regardless of how many parallel circuits it carries.  Since the number
of parallel circuits directly determines both impedance ($N$ circuits in
parallel divide impedance by~$N$) and transfer capacity ($N$ circuits
multiply MVA by~$N$), this omission has a first-order impact on power
flow results.

The pipeline addresses this through two independently calibrated scaling
factors applied to each branch:

\begin{enumerate}[leftmargin=*]
\item \textbf{Topology factor} ($N_T$): models the equivalent number of
  parallel circuits for impedance purposes.
  $R_\text{scaled} = R / N_T$,\;
  $X_\text{scaled} = X / N_T$,\;
  $B_\text{scaled} = B \times N_T$,\;
  $\text{MVA}_\text{scaled} = \text{MVA} \times N_T$.
  Calibrated for AC-OPF convergence (voltage profiles, reactive
  balance, angle stability).

\item \textbf{Capacity factor} ($N_C$): independently scales only the
  thermal rating without altering impedance, calibrated against
  EIA~Electric Power Annual circuit-mile data~\cite{eia_epa2024}:
  $\text{MVA}_\text{final} = \text{MVA}_\text{LUT} \times N_T \times N_C$.
\end{enumerate}

\Cref{tab:topo-factors} lists the single-state calibration values.
Lower-voltage sub-transmission lines (69--161~kV) require the largest
correction because they are smaller, harder to distinguish from
distribution in aerial imagery, and less likely to attract mapper
attention.  EHV lines (230--345~kV) are well-mapped and need minimal
correction.  The 765~kV capacity-factor anomaly ($N_C = 2.0$) reflects
the finding from \cref{sec:voltage-filter} that OSM captures only about
half the 765~kV circuit-miles reported by EIA.

\begin{table}[h!]
\centering\small
\caption{Topology and capacity factors for single-state models.}
\label{tab:topo-factors}
\begin{tabular}{@{}rcc@{}}
\toprule
Voltage (kV) & $N_T$ & $N_C$ \\
\midrule
69  & 3.00 & 1.5 \\
115 & 2.25 & 1.0 \\
138 & 1.75 & 1.0 \\
161 & 1.50 & 1.0 \\
230 & 1.00 & 1.0 \\
345 & 1.00 & 1.0 \\
525 & 1.25 & 1.0 \\
765 & 1.00 & 2.0 \\
\bottomrule
\end{tabular}
\end{table}

When single-state models are merged into multi-state regional models,
each single-state factor is multiplied by an additional per-voltage
scalar: $\times 3$ for the topology factor and $\times 2$ for the
capacity factor (e.g.\ 69~kV rises from $N_T = 3.0$ to~9.0, from
$N_C = 1.5$ to~3.0).  These boosts are necessary because interstate
transit flows -- power routing through one state's corridors to serve
load in another -- demand additional impedance reduction and capacity
headroom that were never needed in isolated single-state models.

The same factors also apply to transformers: impedance is
divided by~$N_T$ (keyed to the low-voltage side) and the MVA rating is
multiplied by the corresponding capacity factor, ensuring that
transformer capacity scales consistently with the lines they connect.

\subsubsection{Generator Cost Curves and Operational Limits}%
\label{sec:gen-costs}

The OPF objective minimizes total generation cost,
$\min \sum_{i \in \mathcal{G}} C_i(P_{g,i})$, which requires a cost
function for every generator.  Since OSM carries no cost information,
these are estimated from fuel type, plant efficiency, and market fuel
prices.

Each generator is assigned a quadratic cost function
$C_i(P_g) = c_2 P_g^2 + c_1 P_g + c_0$, where $c_1$ is the marginal
cost~(\$/MWh) and $c_0$ the no-load cost~(\$/h)~\cite{kirschen2026fundamentals}. Other cost functions can also be used. Cost parameters are
determined through a priority hierarchy:

\begin{enumerate}[leftmargin=*]
\item \textbf{Plant-specific EIA-923 heat rate} (highest priority).
  When the pipeline matches an OSM generator to an EIA-860 plant record
  by name, fuel type, and geographic proximity
  (${\le}\,5$~km)~\cite{eia860_2024}, the plant's actual heat rate
  (BTU/kWh) from EIA-923~\cite{eia923_2024} computes the marginal cost:
  \begin{equation}\label{eq:heatrate}
    c_1 = \frac{\text{Heat Rate} \times \text{Fuel Price}}{1000}
          + \text{VOM}.
  \end{equation}
  A logarithmic size-adjustment curve scales the heat rate, penalizing
  smaller units (which tend to be less efficient) and rewarding larger
  ones, clamped to the range $[0.9,\;1.3]$.  For natural gas generators,
  the fuel price defaults to \$3.50/MMBtu but is overridden at run time
  with the current Henry Hub spot price fetched from the EIA Natural Gas
  API, ensuring that gas-plant costs reflect market conditions at the
  time of analysis.

\item \textbf{Static fuel-type LUT} (fallback).  When EIA matching
  fails, costs are assigned from a default table indexed by
  canonical fuel type (\cref{app:gen-params}), covering 14 fuel
  categories with marginal costs ranging from \$0/MWh (solar, wind) to
  \$90/MWh (diesel peakers).
\end{enumerate}

\noindent
\textbf{Fuel-type normalization.}  OSM and EIA use inconsistent naming
for fuel types (e.g.\ \texttt{natural\_gas}, \texttt{lng},
\texttt{combined\_cycle}, \texttt{ccgt}, and \texttt{gas\_cc} all
represent natural-gas generation).  A shared two-level mapping resolves
this: 48~raw string variants are first normalized to
18~canonical technical types (e.g.\ \texttt{gas}, \texttt{gas\_turbine},
\texttt{solar}), which are then collapsed to 12~display categories
(Solar, Wind, Hydro, Geothermal, Nuclear, Gas, Coal, Oil, Biomass,
Waste, Battery, Unknown).  The display categories define the
\texttt{RENEWABLE\_FUELS} set (Solar, Wind, Hydro, Geothermal) and the
\texttt{ZERO\_MARGINAL\_COST} set (renewables plus Nuclear) used for
merit-order dispatch and decommitment protection.

\begin{figure}[h!]
\centering
\includegraphics[width=\columnwidth]{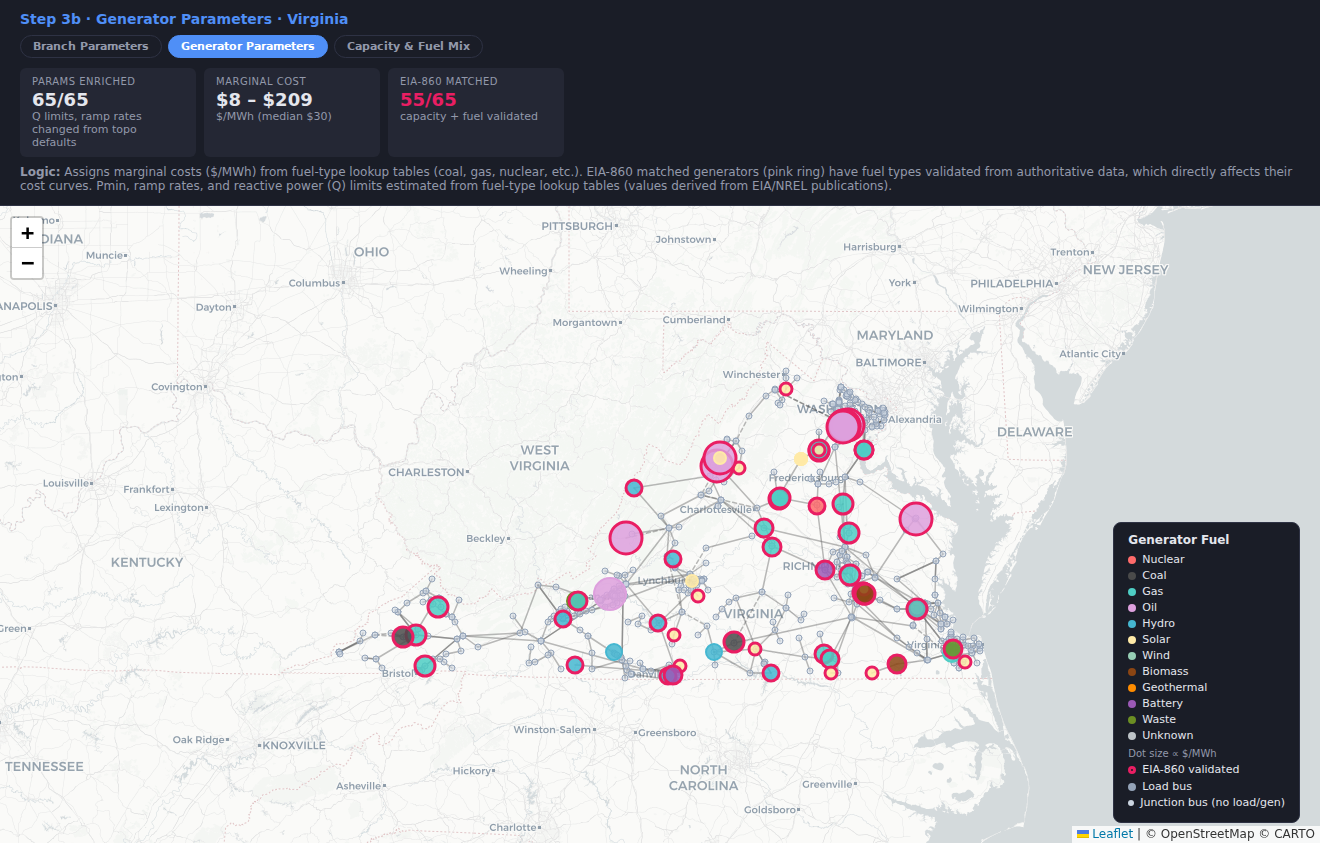}
\caption{Generator parameters for Virginia: 65 generators colored by
  fuel type, sized proportional to marginal cost~(\$/MWh).  Pink rings
  indicate EIA-860--validated plants (55 of 65).  Marginal costs range
  from \$8/MWh~(nuclear) to \$209/MWh~(oil peakers).}
\label{fig:params-va-gencost}
\end{figure}

\noindent
For Virginia, 55 of 65 generators are successfully matched to EIA-860
plant records (\cref{fig:params-va-gencost}), yielding marginal costs of
\$8--\$209/MWh (median~\$19/MWh).

\textbf{Minimum output.}  Thermal generators cannot operate below a
minimum stable output.  The pipeline assigns $P_\text{min}$ as a
fraction of $P_\text{max}$ by fuel type: 50\% for nuclear (must-run
baseload), 30\% for coal, 20\% for gas CCGT, and 0\% for gas turbines,
renewables, and storage.

\subsubsection{Reactive Power and Voltage Limits}\label{sec:reactive}

AC-OPF requires reactive-power limits ($Q_\text{max}$, $Q_\text{min}$)
and voltage bounds for each bus -- parameters that OSM does not provide.
Generator reactive capability is derived from a technology-dependent
rated power factor (PF)
\begin{equation}\label{eq:qmax}
    Q_\text{max} = P_\text{max} \times \tan(\cos^{-1}(PF))
\end{equation}


Synchronous machines (thermal and hydro) use power factors of 0.80--0.90
(nuclear~0.90, coal/gas~0.85, hydro~0.80), providing substantial
reactive support.  Inverter-based resources (solar, wind, battery) are
limited to $\cos\varphi = 0.95$.  The absorption capability
($Q_\text{min}$) is a technology-dependent fraction of~$Q_\text{max}$:
synchronous machines absorb 40--60\% of their rated~$Q_\text{max}$,
while inverter-based solar and battery are symmetric
($Q_\text{min} = -Q_\text{max}$).

\begin{figure}[h!]
\centering
\includegraphics[width=\columnwidth]{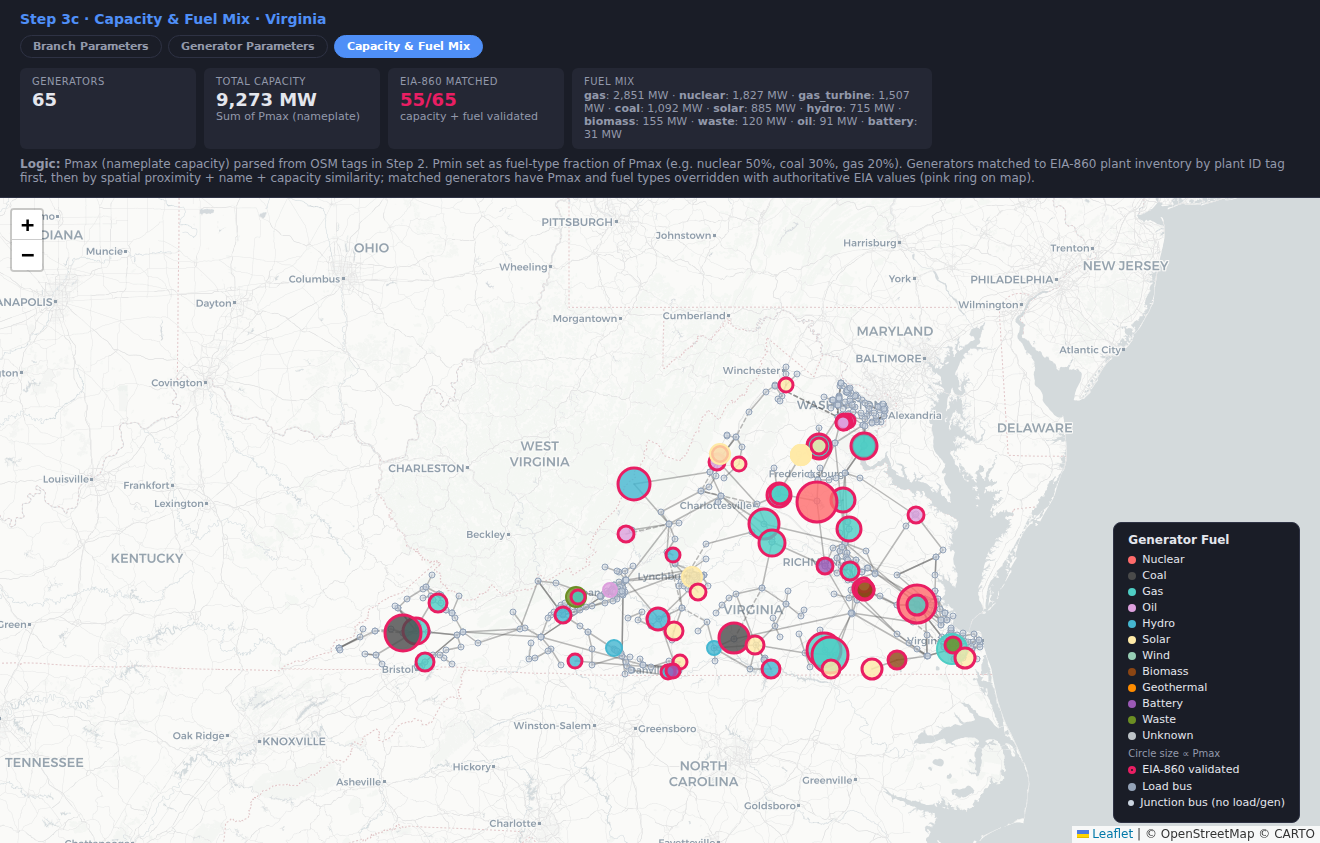}
\caption{Capacity and fuel mix for Virginia: 65 generators totaling
  9{,}273~MW nameplate, sized proportional to~$P_\text{max}$.  Gas
  dominates (2{,}851~MW CCGT $+$ 1{,}507~MW simple-cycle), followed by
  nuclear~(1{,}827~MW), coal~(1{,}092~MW), solar~(885~MW), and
  hydro~(715~MW).}
\label{fig:params-va-gencap}
\end{figure}

Bus voltage magnitude bounds follow standard planning
practice~\cite{glover2012power}: load buses
$V \in [0.95, 1.05]$~pu, generator buses $V \in [0.95, 1.10]$~pu.
Branch angle differences are constrained by voltage class:
$\pm 30$\textdegree{} for EHV ($\ge$100~kV),
$\pm 45$\textdegree{} for subtransmission, and
$\pm 60$\textdegree{} for transformers.
\Cref{fig:params-va-gencap} summarizes the resulting capacity and fuel
mix for Virginia.

\subsection{Demand Allocation (Step 4)}\label{sec:demand}

OSM provides the physical infrastructure; operational data -- how much
power is consumed, where, and when -- must come from public sources.
This stage fetches hourly demand from EIA-930~\cite{eia930_2026},
identifies the Balancing Authorities (BAs) that serve the modeled
region, and distributes load to individual buses using
census-tract population as a spatial proxy.

\subsubsection{Balancing Authority Detection}\label{sec:ba-detect}

\begin{figure}[h!]
\centering
\includegraphics[width=\columnwidth]{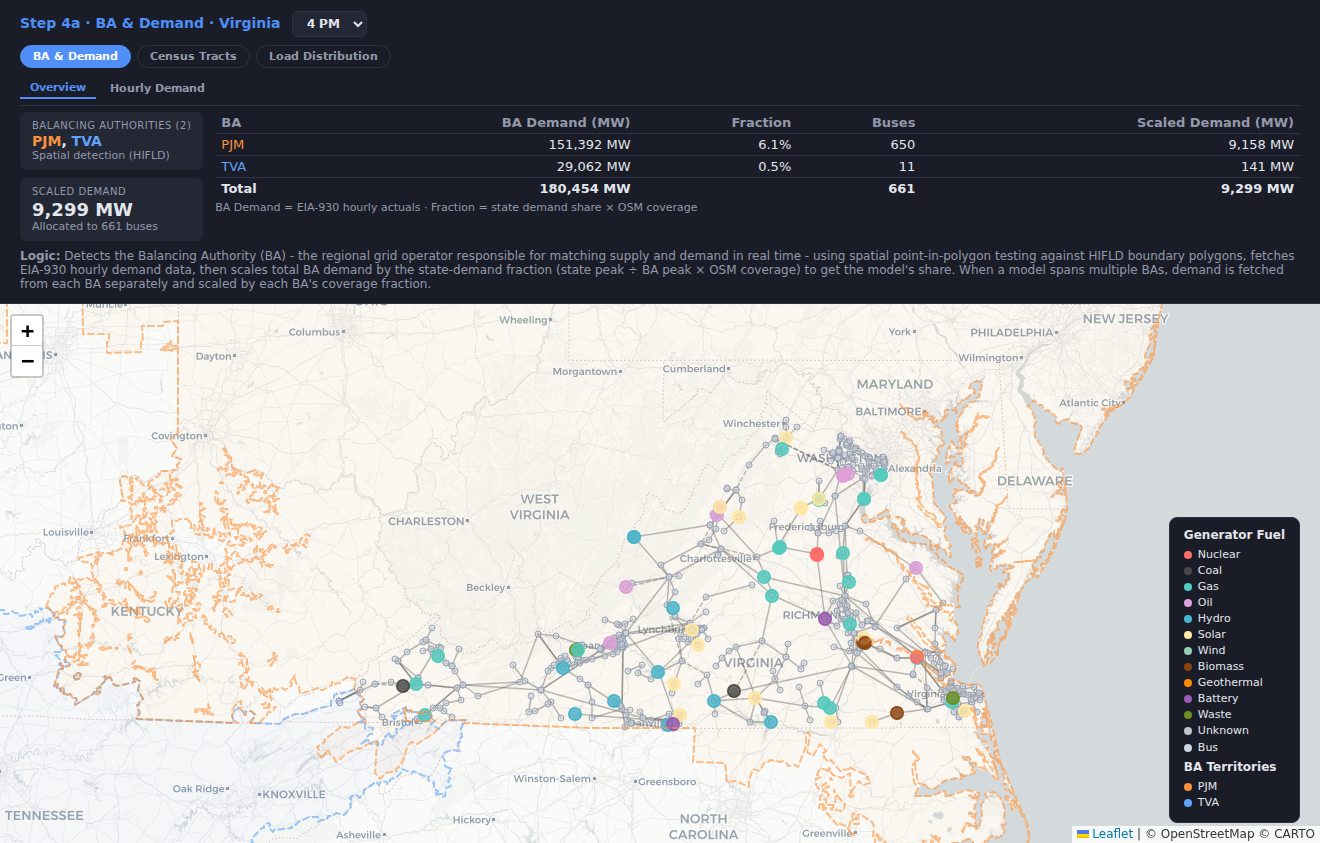}
\caption{BA detection for Virginia at 4~PM: PJM serves 650 buses
  (6.1\% of BA capacity $\to$ 9{,}158~MW); TVA covers 11
  south-western buses (0.5\% $\to$ 141~MW).  Total scaled demand:
  9{,}299~MW.}
\label{fig:demand-va-ba}
\end{figure}

A single state may span multiple BAs.  Virginia, for example, is
predominantly within PJM but includes a small TVA footprint in the
southwest (\cref{fig:demand-va-ba}).  The pipeline assigns each bus to a
BA via point-in-polygon testing against HIFLD BA boundary
polygons~\cite{hifld_ba}.  The BA containing the most buses becomes the
primary BA; secondary BAs are retained when their bus share exceeds~1\%
and at least one generator is present.  Sub-BAs (e.g.\ Duke Energy
Progress, PacifiCorp~West) that do not publish standalone EIA-930 demand
data are resolved to their parent BA via a mapping table -- for instance,
CPLE $\to$ DUK and PACW $\to$ PACE -- ensuring demand data is always
available.

\textbf{Demand scaling.}  For each detected BA, hourly demand is fetched
from EIA-930 and scaled by a \emph{regional fraction}~$f$ that
estimates what share of the BA's load the model represents:
\begin{equation}\label{eq:ba-scale}
  D_\text{model} = D_\text{BA} \times f.
\end{equation}
The fraction is computed differently depending on the model scope,
with three cases handled by a tiered strategy:

\begin{enumerate}[leftmargin=*]
\item \textbf{Single-state, single-BA}
  (e.g.\ ERCOT/Texas, CAISO/California).
  When the BA serves exactly the modeled state, the fraction is
  \begin{equation}\label{eq:frac-single}
    f = \frac{D_\text{state}^{\,\text{peak}}}{D_\text{BA}^{\,\text{peak}}},
  \end{equation}
  where $D_\text{state}^{\,\text{peak}}$ is the state's summer peak
  demand from EIA-861~\cite{eia861_2024} and
  $D_\text{BA}^{\,\text{peak}}$ is the BA's peak from EIA-930.
  For a true single-state BA this ratio is $\approx$1.0, requiring no
  approximation.

\item \textbf{Single-state, multi-BA}
  (e.g.\ Virginia spanning PJM and TVA).
  Buses are partitioned by their assigned BA\@.  For each BA~$k$, the
  fraction is the state peak scaled by the BA's bus share and by
  the model's \emph{OSM coverage} -- the ratio of OSM-captured
  generation capacity to the state's peak demand:
  \begin{equation}\label{eq:frac-multi-ba}
    f_k = \frac{D_\text{state}^{\,\text{peak}} \;\times\; b_k}
               {D_{k}^{\,\text{peak}}} \;\times\; c_\text{osm},
    \qquad
    c_\text{osm} = \min\!\Bigl(1,\;
      \frac{\sum P_\text{max}^{\,\text{osm}}}
           {D_\text{state}^{\,\text{peak}}}\Bigr),
  \end{equation}
  where $b_k$ is the fraction of model buses assigned to BA~$k$.
  The coverage cap prevents the model from claiming more load than its
  captured infrastructure can plausibly serve.  For Virginia,
  $c_\text{osm} \approx 0.64$, reflecting that OSM captures roughly
  two-thirds of the generation needed to meet the state's peak.

\item \textbf{Multi-state region}
  (e.g.\ PJM 14~states, Western 11~states).
  For each BA that overlaps the region, the fraction is the sum of
  state peaks for the overlapping states, divided by the BA peak. 
  For single-state BAs within the region, the fraction is used
  directly (analogous to case~1); for multi-state BAs, it is
  multiplied by a capacity-based coverage ratio within that BA:
  \begin{equation}\label{eq:frac-region}
    f_k = \frac{\sum_{s \in S_k} D_s^{\,\text{peak}}}
               {D_k^{\,\text{peak}}}
    \times
    \begin{cases}
      1 & \text{if BA $k$ is single-state,} \\[4pt]
      \min\!\bigl(1,\;
        P_\text{max}^{\,\text{model},k} / P_\text{max}^{\,\text{BA},k}\bigr)
        & \text{otherwise.}
    \end{cases}
  \end{equation}
  Here $S_k$ is the set of modeled states served by BA~$k$, and
  $P_\text{max}^{\,\text{model},k}$ is the OSM-captured generation
  capacity within BA~$k$.  The coverage multiplier for multi-state BAs
  prevents over-allocation when the model covers only a subset of
  the BA's geographic footprint
  (e.g.\ MISO spans states both inside and outside the PJM model).
\end{enumerate}

Demand is always fixed before any EIA generator injection
(\cref{sec:dispatch}), preventing a circular feedback between demand and
capacity.

For Virginia at 4~PM, PJM reports 151{,}392~MW; the state-demand
fraction (6.1\%) yields 9{,}158~MW.  TVA contributes 141~MW from
11~buses, bringing the total scaled demand to 9{,}299~MW
(\cref{fig:demand-va-ba}).

\subsubsection{Census-Based Load Allocation}\label{sec:census-alloc}

When the model spans multiple BAs, buses are partitioned by their
assigned BA and demand is allocated independently within each partition.
Within each partition, total demand is distributed to individual buses
using US Census tract population as a spatial proxy for electricity
consumption.  Residential
and commercial loads are the dominant demand components and both track
population; this approach is consistent with methods used in
PyPSA-Eur~\cite{horsch2018}.

The pipeline downloads TIGER/Line census-tract boundaries and
ACS~5-Year population estimates from the Census Bureau API, then
performs a spatial join: each bus is assigned to the tract it falls
within (with a nearest-neighbour fallback for unmatched buses).  The
bus's share of total demand is proportional to the population in its
tract:
\begin{equation}\label{eq:census-load}
  P_{d,i} = D_\text{model}
    \times \frac{\text{pop}_i}{\sum_j \text{pop}_j}.
\end{equation}
The reactive component
$Q_{d,i} = P_{d,i} \times \tan(\arccos 0.92)$ uses a load power factor
of~0.92.  For Virginia, 2{,}198 census tracts are matched to 661~buses.

\subsubsection{Generation Dispatch}\label{sec:dispatch}

Before solving the OPF problem to optimize generation, it is useful to compute a rough estimate of the generation levels to gain some intuitive understanding and serve as initialization for the optimization problem. Our pipeline sets initial generator outputs that balance total generation with total load.  A 3\% loss factor is applied ($D_\text{gross} = D_\text{model} \times 1.03$) to account for resistive and transformer losses.

Generators are dispatched in \emph{merit order}: they are sorted by
marginal cost and committed cheapest-first until cumulative output meets
$D_\text{gross}$.

\noindent
\textbf{Renewable derating.}  Solar and wind are the only fuel types
treated as intermittent; all others (including hydro and nuclear) are
fully dispatchable.  Each intermittent generator's $P_{\max}$ is
multiplied by an hour- and season-dependent capacity factor drawn from
idealized profiles covering three seasons (summer: Jun--Aug; winter:
Dec--Feb; spring/fall: all other months).  Solar capacity factors range
from~0 overnight to a summer-noon peak of~0.95, with winter noon
reaching only~0.70 and spring/fall~0.85.  By 4~PM, the summer
solar factor drops to~0.52, so Virginia's 885~MW of nameplate solar is
derated to~$885 \times 0.52 \approx 460$~MW.  Wind capacity factors are
more uniform across the day: summer values range from~0.20 (pre-dawn) to
0.60 (afternoon), while winter profiles are higher and flatter
(0.35--0.58).  At night, wind capacity factors remain 0.25--0.42,
making wind a significant contributor to off-peak generation.

\begin{figure}[h!]
\centering
\includegraphics[width=\columnwidth]{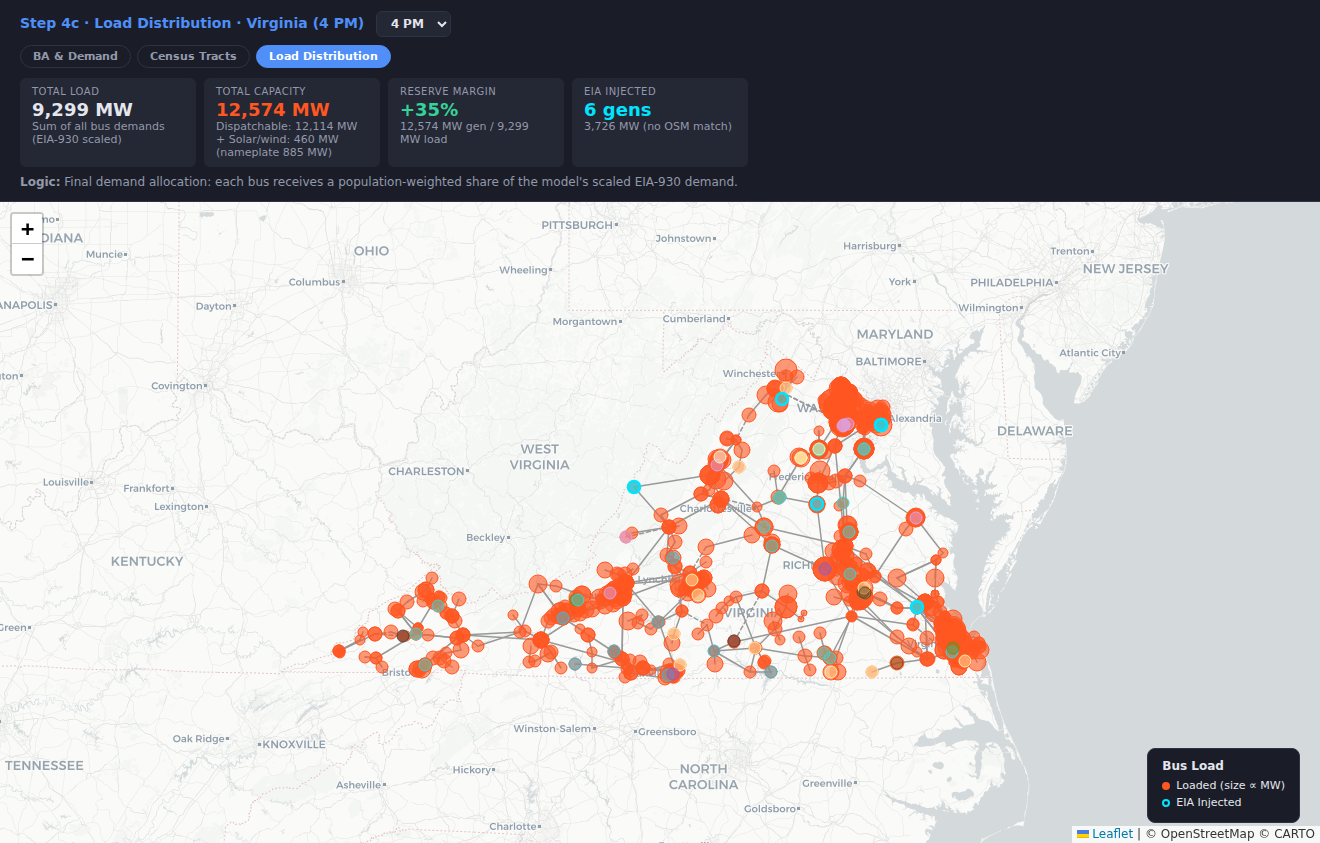}
\caption{Final load distribution for Virginia at 4~PM: 9{,}299~MW
  allocated to 661~buses (sized proportional to load), 12{,}574~MW
  available capacity ($+$35\% reserve margin).  Six EIA-injected
  generators (cyan rings) had no OSM match.}
\label{fig:demand-va-loads}
\end{figure}

\textbf{EIA generator injection.}  When the model's total generation
capacity falls below a 30\% reserve margin above scaled demand, the
pipeline injects additional generators from EIA-860 that were not found
in OSM.  Unmatched EIA plants are sorted by capacity descending and
assigned to the nearest bus with available connection slots
($\le$50~km), each bus limited by its branch degree.
Injection continues until the 30\% reserve margin floor is met. 
For Virginia, six generators are injected this way: two nuclear plants
(North Anna~980~MW, Surry~848~MW), two gas units at Chalk Point
(659~MW each), Warren County gas~(580~MW), and Bath County
hydro~(1~MW)
(\cref{fig:demand-va-loads}).

\textbf{Slack bus.}  After dispatch, the slack bus is reassigned to the
largest dispatchable (non-renewable) generator, ensuring sufficient
headroom to absorb power imbalances and the feasibility of the OPF problem.

\subsection{Optimal Power Flow (Step 5)}\label{sec:opf}

The final pipeline step solves an Optimal Power Flow (OPF) problem on the
model produced by Steps~2--4.  The implementation uses
PowerModels.jl~\cite{coffrin2018} interfacing with the Ipopt
interior-point solver~\cite{wachter2006}.

\subsubsection{Input Format and Solver Configuration}\label{sec:opf-input}

The solver receives a single JSON file in MATPOWER/PowerModels format
containing buses, branches (AC lines and transformers), DC lines,
generators, loads, and shunt elements.  All values are expressed in
per-unit on a 100~MVA base (\cref{app:perunit}).  Cost coefficients,
voltage bounds, angle limits, and thermal ratings are included.

Ipopt is configured with a primary convergence tolerance of~$10^{-4}$,
an acceptable (relaxed) tolerance of~$10^{-2}$, and a maximum of
10{,}000 iterations.
Solutions meeting the strict tolerance receive
\texttt{LOCALLY\_SOLVED} status; those converging only within the
relaxed tolerance receive \texttt{ALMOST\_LOCALLY} \texttt{\_SOLVED} and are
accepted as qualified successes.

\subsubsection{DC and AC Formulations}\label{sec:opf-formulation}

\textbf{DC-OPF.}  The linear DC approximation is solved first as a
screening tool and warm-start seed.  It minimizes the same cost
objective but with a simplified constraint set.  DC-OPF solves
in under one second for networks up to $\sim$5{,}000 buses.  Like the
AC formulation, it is attempted with progressive relaxation.

\textbf{AC-OPF.}  The full nonlinear formulation includes voltage
magnitudes, angles, and both real and reactive power at every bus.  The
DC-OPF solution provides initial values for bus voltage angles and
generator dispatches; these are passed to the AC solver as starting
points, dramatically improving convergence compared to a flat-voltage
cold start.

\subsubsection{Progressive Relaxation}\label{sec:relaxation}

Standard OPF assumes a well-characterized network with precise
parameters.  OSM-derived models violate this assumption: impedances are
estimated from LUTs, parallel circuits are approximated by
scaling factors, and demand allocation is a spatial heuristic.  The
result is that some models are infeasible under strict constraints -- not
because the underlying grid is infeasible, but because the model's
parameters are imprecise.

Rather than requiring manual tuning per state, the pipeline
automatically loosens constraints through a sequence of \emph{relaxation
levels} until convergence is achieved.  Six cumulative levels (L0--L5)
are defined, plus an AC-specific base layer (AC1) (\cref{tab:relaxation}):

\begin{table}[h!]
\centering\small
\caption{Progressive relaxation levels.  Each level is cumulative,
  including all relaxations from previous levels.}
\label{tab:relaxation}
\begin{tabular}{@{}clcccccc@{}}
\toprule
Level & Name & Angle & Thermal & $V_\text{min}/V_\text{max}$ &
  $Q$ factor & Load cap & $P_\text{min}$ \\
\midrule
L0 & Strict       & default & $1.0\times$ & default & $1.0\times$ &
  100\% & default \\
L1 & Widen angles & $\pm60^{\circ}$ & $1.0\times$ & default &
  $1.0\times$ & 100\% & default \\
L2 & Thermal headroom & $\pm60^{\circ}$ & $1.2\times$ & default &
  $1.0\times$ & 100\% & default \\
L3 & Aggressive   & $\pm90^{\circ}$ & $1.5\times$ & default &
  $1.0\times$ & 100\% & $\times 0.5$ \\
L4 & Load shedding & $\pm90^{\circ}$ & $1.5\times$ & default &
  $1.0\times$ & 70\% & 0 \\
L5 & Full relaxation & $\pm90^{\circ}$ & $\infty$ &
  [0.85,\,1.15] & $2.0\times$ & 70\% & 0 \\
\midrule
AC1 & Voltage+$Q$ relax &  -  &  -  & [0.90,\,1.10] &
  $1.5\times$ &  -  &  -  \\
\bottomrule
\end{tabular}
\end{table}

Each level targets a specific class of infeasibility:
L1 addresses angle-limit violations on long, high-impedance branches;
L2 relieves thermal congestion from underestimated branch capacity;
L3 reduces minimum-generation constraints on heavily loaded networks;
L4 curtails load to 70\% of total generation capacity when the
network cannot deliver full demand;
L5 removes essentially all binding constraints as a last resort.
AC1 widens voltage bounds and increases reactive power limits, and is
activated as a persistent base layer for all AC-OPF attempts because
OSM-derived models frequently have reactive power imbalances from
approximate line-charging parameters.

\textbf{Impedance consistency.}  Before each OPF solve, a preprocessing
pass ensures that every branch satisfies
$\text{rate\_a} \times x \le \pi/2$, so that the DC power-flow
constraint $P \le \text{rate\_a}$ remains compatible with the angle
bounds.  This check is applied at L0 and within each relaxation level
(L1--L4); at L5, where thermal limits are removed
($\text{rate\_a} = 10^{6}$), it is skipped.  The fix eliminates
spurious DC-OPF infeasibility on aggregated parallel circuits whose
combined reactance, after topology-factor scaling, would otherwise
exceed the feasible angle range.

\textbf{Generator decommitment.}  Before any OPF solve, a preprocessing step
checks whether total minimum generation ($\sum P_{\min}$) exceeds total
demand.  If so, the most expensive dispatchable generators have their
$P_{\min}$ set to zero, allowing the OPF to dispatch them at zero
output while keeping them grid-connected for reactive power support.
Nuclear and renewable generators are protected from decommitment.  This
step reduces the need for the $P_{\min}$ relaxation at~L3/L4.

\textbf{Convergence strategy.}  The solver proceeds in three stages:
\begin{enumerate}
\item \emph{DC progressive relaxation.}  Attempt DC-OPF at L0; if
  infeasible, escalate through L1--L5.
\item \emph{Reactive shunt injection.}
  Shunt devices (capacitor banks and reactors) provide localized
  reactive-power compensation in real grids, but OSM carries no
  information about them.  To fill this gap, the solver uses the
  DC~dispatch to estimate per-bus reactive imbalance:
  for each bus, reactive demand and estimated branch losses
  ($Q_{\text{loss}} \approx P_{ij}^{2} x_{ij}$, computed from the
  DC~power flows and branch reactances) are summed against available
  generator reactive capability and line-charging injection.  Where
  the deficit (or surplus from excess line charging on long HV~lines)
  exceeds a 15\% margin, a compensating shunt capacitor (or reactor)
  is inserted.  This dispatch-aware pre-conditioning targets
  compensation to buses that actually need it under the solved
  operating point, rather than relying on heuristic placement, and
  prevents reactive-power shortfall from being the sole cause of AC
  infeasibility.
  Because every bus with an unmet reactive deficit receives a shunt,
  the resulting coverage is high ($\sim$90\% of buses) compared to
  utility-grade reference models such as PGLib-OPF, where discrete
  physical devices appear at 5--15\% of buses.
  This higher density reflects a modeling choice: our shunts act as
  proxy for all reactive resources absent from OSM,
  rather than representing individually cataloged equipment as in
  reference models.
\item \emph{AC progressive relaxation.}  Attempt AC-OPF at L0 with the
  DC warm-start and injected shunts.  If infeasible, activate AC1,
  then escalate through L1--L5 (each with AC1 as the persistent base
  layer).  Stop at first convergence.
\end{enumerate}
Each AC relaxation level is attempted as a separate subprocess with a
1{,}800-second timeout, ensuring that a stuck Ipopt solve does not block the
pipeline.  Every subprocess starts from the same DC-warm-started model.

The relaxation level achieved is reported alongside all results, serving
as both a convergence indicator and a model quality metric: a state
solving at L0 has a well-characterized network; one requiring L4--L5
has significant data gaps.

\subsubsection{Virginia Example}\label{sec:opf-virginia}

For Virginia at 4~PM, both formulations converge at L0 (strict) with
\texttt{LOCALLY\_SOLVED} status.  DC-OPF produces a total cost of
\$186{,}589/hr (\$20.1/MWh); AC-OPF yields \$188{,}339/hr (\$20.3/MWh),
a~0.9\% AC premium attributable to resistive losses and reactive power
constraints.  Total generation is 9{,}392~MW against 9{,}299~MW of
load, with 93~MW of losses (1.0\%).  AC-OPF solves in 21~seconds.

\begin{figure}[h!]
\centering
\includegraphics[width=\columnwidth]{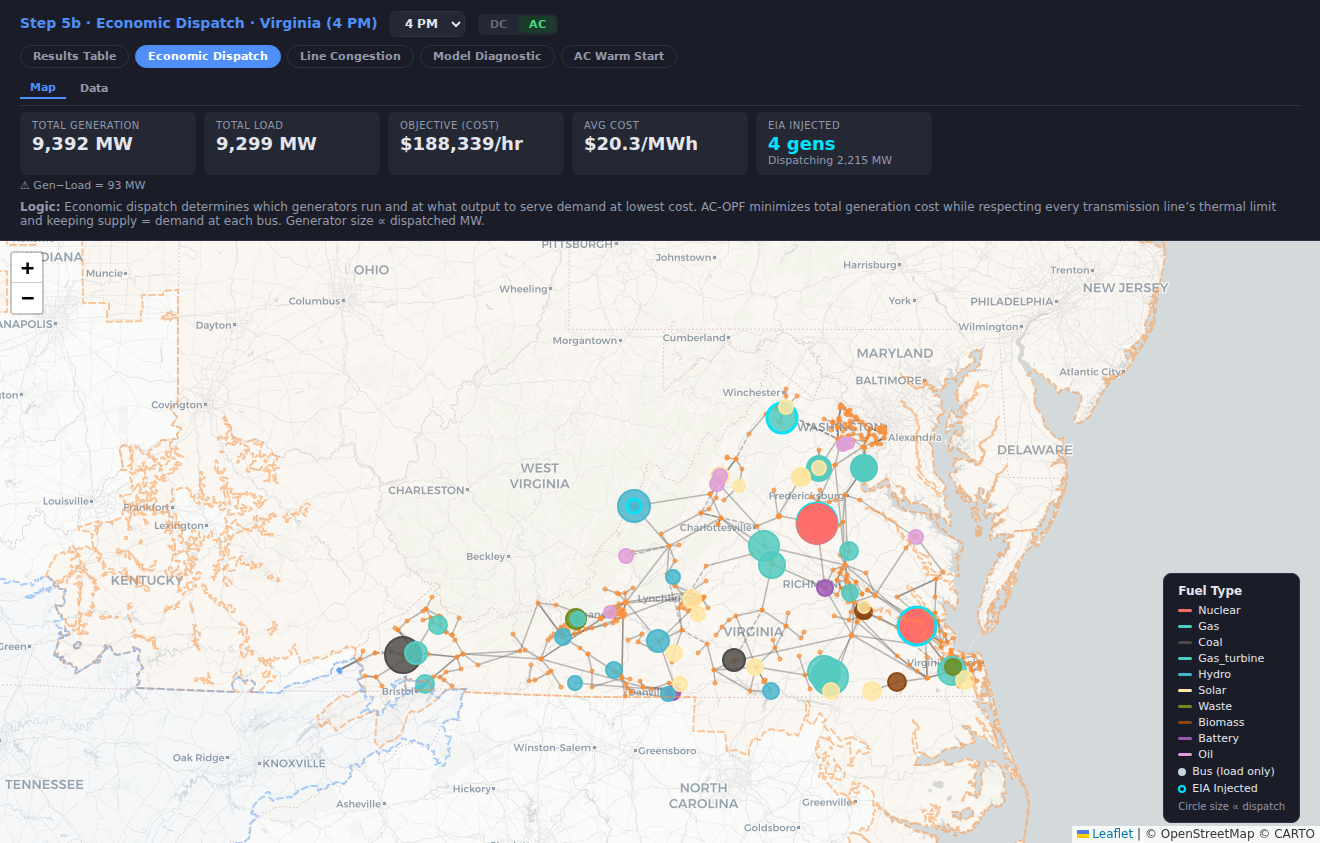}
\caption{AC-OPF economic dispatch for Virginia at 4~PM: 9{,}392~MW
  total generation across 71~generators, \$20.3/MWh average cost.
  Generator circle size $\propto$ dispatched~MW; color indicates
  fuel type.}
\label{fig:opf-va-dispatch}
\end{figure}

\begin{figure}[h!]
\centering
\includegraphics[width=\columnwidth]{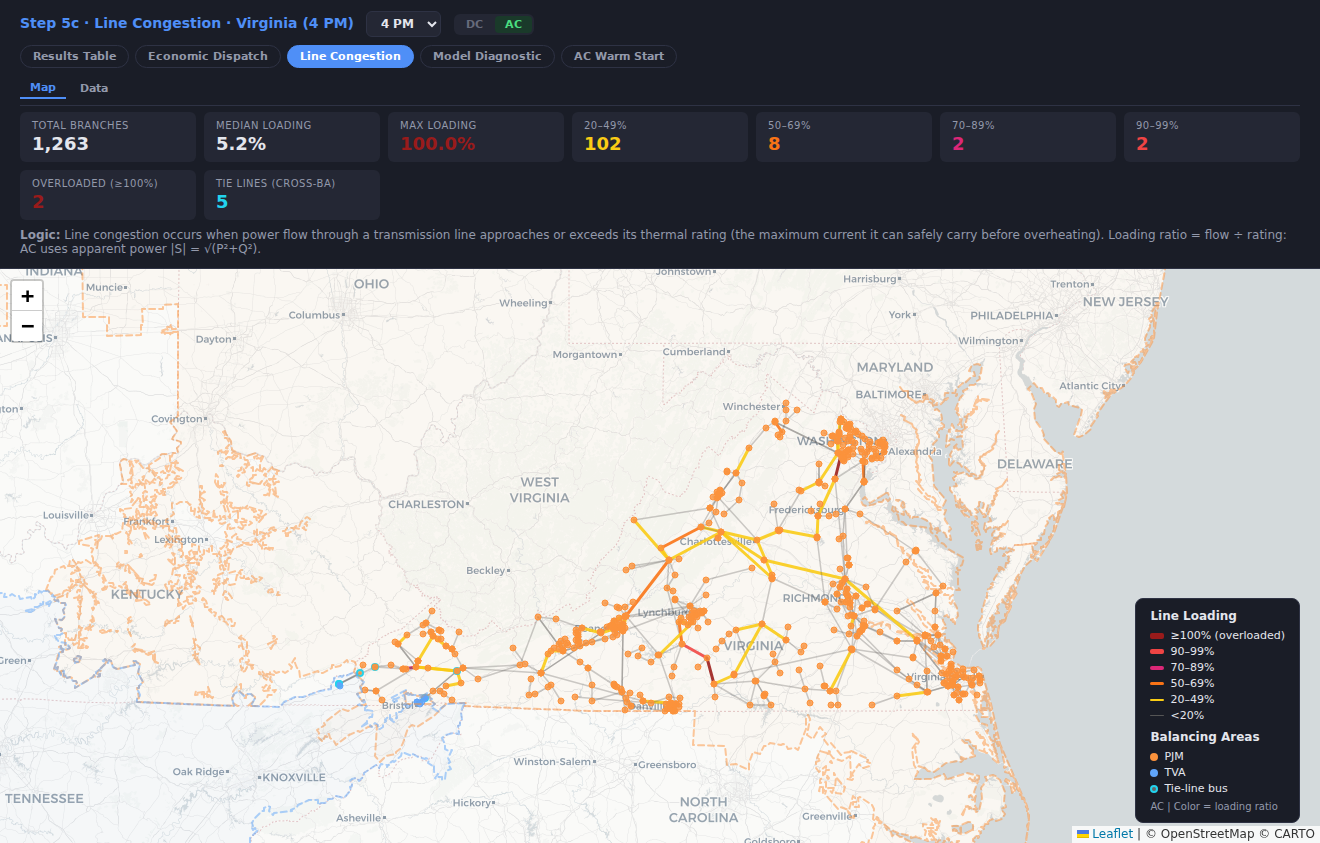}
\caption{AC-OPF line congestion for Virginia at 4~PM: 1{,}263
  branches colored by loading ratio (flow\,/\,thermal rating).
  Median loading 5.2\%; two branches reach 100\%.  Congestion
  concentrates in the northern Virginia--Washington corridor.}
\label{fig:opf-va-congestion}
\end{figure}

\Cref{fig:opf-va-dispatch} shows the economic dispatch map: generator
circles are sized proportional to dispatched MW and colored by fuel
type.  Nuclear (red) and gas (teal/green) dominate, with four
of six EIA-injected generators dispatching.  The dispatch is dominated by
nuclear (39\%), gas (31\%), coal (8\%), gas turbine (7\%), and
hydro (7\%), which together account for over 92\% of generation.
\Cref{fig:opf-va-congestion}
shows branch loading ratios: most lines operate below 20\% of their
thermal rating (grey), while a cluster of branches in the northern
Virginia corridor and around Lynchburg reach 50--90\% loading
(orange/pink), consistent with the high population density in those
areas.

\subsubsection{Multi-State Regional Models}\label{sec:multi-state}

The pipeline supports merging per-state OSM downloads into multi-state
regional models, enabling analysis of interstate power flows and
regional dispatch patterns invisible in single-state models.  Per-state
GeoJSON files are combined, deduplicating features that appear in
overlapping bounding boxes by their OSM identifier.  The merged dataset
then passes through the same pipeline (Steps~2--5).

\textbf{Transit flows and internal topology boost.}  Merging
introduces interstate transit flows (\cref{sec:topology-factors}),
which overload OSM's single-circuit corridors.  The multi-state
topology and capacity factors described in the same section address
this by multiplying the single-state factors by~$\times 3$ (impedance)
and $\times 2$ (capacity).

\section{\texorpdfstring{Results}{Results}}\label{sec:results}

Every model is solved at both 4~AM (off-peak) and
4~PM (peak) using the EIA-930 demand profile for
July~15, 2024 (a summer weekday).  Unless stated otherwise, figures
refer to the 4~PM (peak) snapshot.

\subsection{Single-State Models}\label{sec:results-single}

The pipeline was run for all 48 contiguous US states.  Every state
achieves Ipopt's \texttt{LOCALLY\_SOLVED} status for both DC-OPF and
AC-OPF -- 96 out of 96 solves converge successfully.  For DC-OPF, 45~states solve
at~L0; Illinois and New York require~L5, and Utah~L3.
For AC-OPF, 42~states solve at~L0 and California requires only the
lightweight AC1 base layer (voltage/reactive relaxation), while
New Mexico and West Virginia need~L2, Utah~L3, and Illinois and
New York~L5.
\Cref{tab:single-state} lists the full results.

\begin{table}[h!]
\centering\footnotesize
\caption{Single-state AC-OPF and DC-OPF results at 4~PM (peak), 
  sorted by Load (MW) from largest to smallest. Negative loss percentages indicate load 
  shedding under high relaxation.}
\label{tab:single-state}
\setlength{\tabcolsep}{3.5pt}
\begin{tabular}{@{}lrrrrrcrrr@{}}
\toprule
State & Load & Buses & Gens & Br & Rlx &
  AC & DC & Loss & AC Time \\
 & (MW) & & & & & (\$/MWh) & (\$/MWh) & (\%) & (s) \\
\midrule
Texas           &   74{,}049 &  3{,}889 & 509 & 6{,}852 & L0  &  15.2 &  14.7 &     1.5 &    50 \\
California      &   31{,}806 &      769 & 323 & 1{,}582 & AC1 &  18.5 &  17.3 &     1.3 &    99 \\
New York        &   27{,}746 &      626 & 645 & 1{,}157 & L5  &  26.0 &  25.4 &     0.9 &   320 \\
Florida         &   27{,}522 &  1{,}773 & 178 & 3{,}161 & L0  &  26.9 &  26.4 &     1.1 &    29 \\
Illinois        &   21{,}177 &      770 &  90 & 1{,}520 & L5  &  15.8 &  14.6 & $-$2.5 &   379 \\
Pennsylvania    &   17{,}070 &      910 & 107 & 1{,}749 & L0  &  22.2 &  22.0 &     0.7 &    21 \\
Ohio            &   16{,}935 &      784 &  96 & 1{,}513 & L0  &  24.9 &  24.4 &     1.4 &    21 \\
Georgia         &   12{,}924 &  1{,}240 & 103 & 2{,}079 & L0  &  18.4 &  18.2 &     0.7 &    23 \\
North Carolina  &   11{,}537 &      704 & 110 & 1{,}509 & L0  &  20.6 &  20.3 &     0.8 &    21 \\
Tennessee       &    9{,}803 &      603 &  65 & 1{,}122 & L0  &  15.8 &  15.5 &     1.2 &    20 \\
Michigan        &    9{,}758 &      607 &  68 & 1{,}131 & L0  &  22.5 &  21.9 &     1.5 &    20 \\
Colorado        &    9{,}559 &      651 &  99 & 1{,}146 & L0  &  12.9 &  12.4 &     1.5 &    20 \\
Louisiana       &    9{,}552 &      571 &  48 &     999 & L0  &  26.6 &  26.4 &     0.6 &    19 \\
Alabama         &    9{,}501 &      494 &  55 &     864 & L0  &  20.9 &  20.6 &     0.9 &    20 \\
Virginia        &    9{,}299 &      661 &  71 & 1{,}263 & L0  &  20.3 &  20.1 &     1.0 &    21 \\
Indiana         &    9{,}204 &  1{,}271 &  97 & 2{,}316 & L0  &  26.8 &  26.4 &     0.9 &    24 \\
Oklahoma        &    9{,}058 &      906 &  64 & 1{,}460 & L0  &  23.1 &  22.6 &     1.0 &    20 \\
Nevada          &    8{,}672 &      296 & 106 &     614 & L0  &  16.0 &  15.8 &     0.4 &    18 \\
Missouri        &    8{,}645 &  1{,}033 &  65 & 1{,}753 & L0  &  29.8 &  29.4 &     1.3 &    22 \\
Arizona         &    7{,}822 &      730 &  78 & 1{,}356 & L0  &  16.8 &  16.7 &     0.5 &    21 \\
Arkansas        &    7{,}726 &      778 &  69 & 1{,}195 & L0  &  20.0 &  19.7 &     1.0 &    20 \\
Utah            &    7{,}364 &      244 & 240 &     419 & L3  &  21.7 &  20.9 &     2.1 &   219 \\
South Carolina  &    7{,}181 &      588 &  67 & 1{,}291 & L0  &  16.3 &  16.0 &     1.0 &    20 \\
Minnesota       &    7{,}131 &      718 &  97 & 1{,}297 & L0  &  22.0 &  21.6 &     1.1 &    21 \\
Mississippi     &    6{,}782 &      528 &  39 &     902 & L0  &  23.2 &  22.8 &     1.2 &    19 \\
Wisconsin       &    6{,}648 &      776 &  65 & 1{,}427 & L0  &  26.6 &  26.0 &     1.5 &    21 \\
Kansas          &    6{,}371 &      653 &  44 & 1{,}071 & L0  &  25.1 &  24.7 &     1.1 &    20 \\
Iowa            &    6{,}315 &      832 &  61 & 1{,}488 & L0  &  22.4 &  22.0 &     1.0 &    21 \\
New Jersey      &    6{,}210 &      258 &  57 &     580 & L0  &  14.6 &  14.4 &     1.0 &    19 \\
Massachusetts   &    5{,}623 &      222 &  99 &     426 & L0  &  22.7 &  21.7 &     0.7 &    18 \\
Nebraska        &    5{,}006 &      466 &  46 &     865 & L0  &  24.3 &  23.8 &     1.0 &    20 \\
Connecticut     &    4{,}994 &      156 &  41 &     281 & L0  &  23.2 &  22.9 &     0.7 &    18 \\
West Virginia   &    4{,}518 &      166 &  13 &     281 & L2  &  33.2 &  32.8 &     0.9 &    75 \\
Kentucky        &    4{,}113 &      701 &  34 & 1{,}327 & L0  &  29.5 &  29.0 &     1.4 &    20 \\
Maryland        &    3{,}749 &      165 &  27 &     385 & L0  &  24.6 &  24.4 &     0.5 &    18 \\
Wyoming         &    3{,}485 &      192 &  26 &     314 & L0  &  30.6 &  29.9 &     2.0 &    18 \\
Oregon          &    3{,}017 &      341 &  61 &     624 & L0  &  13.6 &  13.5 &     0.3 &    19 \\
Idaho           &    2{,}780 &      179 & 149 &     328 & L0  &   4.4 &   4.3 &     1.1 &    19 \\
New Mexico      &    2{,}268 &      150 &  24 &     252 & L2  &  19.1 &  18.6 &     1.4 &    72 \\
Washington      &    2{,}216 &      724 &  66 & 1{,}211 & L0  &  11.3 &  11.2 &     0.2 &    19 \\
New Hampshire   &    2{,}062 &       81 &  19 &     125 & L0  &  18.2 &  17.8 &     1.5 &    17 \\
North Dakota    &    1{,}797 &      312 &  10 &     561 & L0  &  31.9 &  31.1 &     1.8 &    18 \\
Montana         &    1{,}778 &      382 &  26 &     771 & L0  &  21.6 &  19.0 &     7.1 &    19 \\
South Dakota    &    1{,}538 &      252 &  13 &     489 & L0  &  23.2 &  22.5 &     1.8 &    19 \\
Delaware        &    1{,}506 &       33 &   9 &      77 & L0  &  18.3 &  18.1 &     0.3 &    17 \\
Vermont         &        937 &       40 &  27 &      58 & L0  &   2.6 &   2.6 &     0.4 &    16 \\
Maine           &        908 &       68 &  18 &     107 & L0  &  19.9 &  19.4 &     1.3 &    17 \\
Rhode Island    &        225 &       11 &   4 &      17 & L0  & 104.1 & 103.7 &     0.3 &    17 \\
\bottomrule
\end{tabular}
\end{table}

Several patterns emerge from the results:

\textbf{Relaxation distribution.}  For DC-OPF, 45 of 48 states solve
at~L0 (94\%); the three exceptions (Illinois~L5, New York~L5,
Utah~L3) are among the states that require elevated AC relaxation.
For AC-OPF, 42~states solve at~L0 (88\%) and California requires only
the AC1 base layer (voltage/reactive relaxation).  New Mexico~(L2) and
West Virginia~(L2) need modest thermal headroom; Utah~(L3) requires
aggressive relaxation; Illinois~(L5) and New York~(L5) require full
relaxation.  Illinois exhibits negative losses ($-$2.5\%), indicating
load shedding to reach feasibility -- a sign that the OSM topology is
too sparse to carry the allocated demand.

\begin{figure}[h!]
\centering
\includegraphics[width=\textwidth]{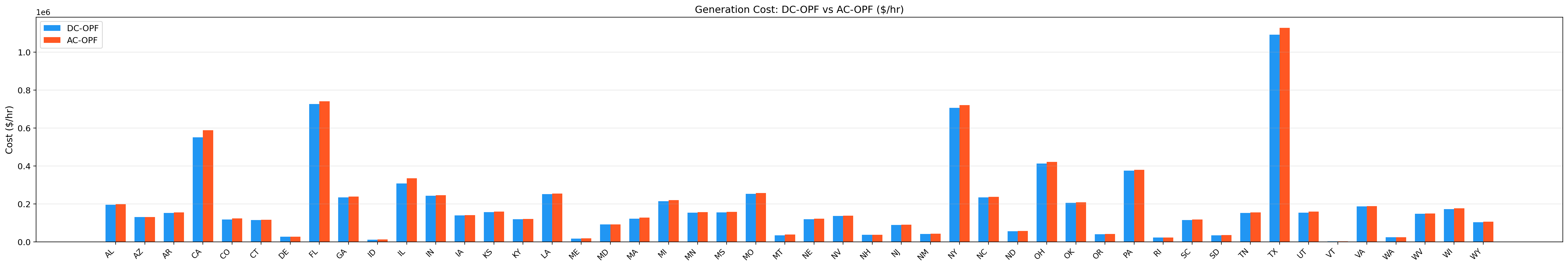}
\caption{DC-OPF vs.\ AC-OPF generation cost (\$/hr) for all
  48~contiguous states at 4~PM (peak).  Bar heights are nearly
  identical for most states; the visible difference in a few cases
  reflects resistive losses captured only by the AC formulation.}
\label{fig:dc-vs-ac-cost}
\end{figure}

\textbf{Cost patterns.} We report the average generation cost,
defined as $\bar{c} = C_{\mathrm{obj}} / P_{\mathrm{load}}$, where
$C_{\mathrm{obj}}$~(\$/hr) is the OPF objective value and
$P_{\mathrm{load}}$~(MW) is the total system load.
The median average AC-OPF cost across states is \$22.1/MWh.  States
with abundant hydropower (Washington~\$11.3, Vermont~\$2.6,
Idaho~\$4.4, Oregon~\$13.6) or wind (Colorado~\$12.9) have the lowest
average costs because these resources have near-zero marginal costs.  States
reliant on coal exhibit higher costs (Kentucky~\$29.5,
West Virginia~\$33.2), as do states with limited local generation that
must import through congested corridors (Rhode Island~\$104.1).  Rhode
Island is an outlier explained by its tiny grid (11~buses,
4~generators) with minimal local generation and high reliance on
expensive peaking units.
\Cref{fig:dc-vs-ac-cost} compares the absolute dispatch cost
for every state under both formulations

\begin{figure}[h!]
\centering
\includegraphics[width=\textwidth]{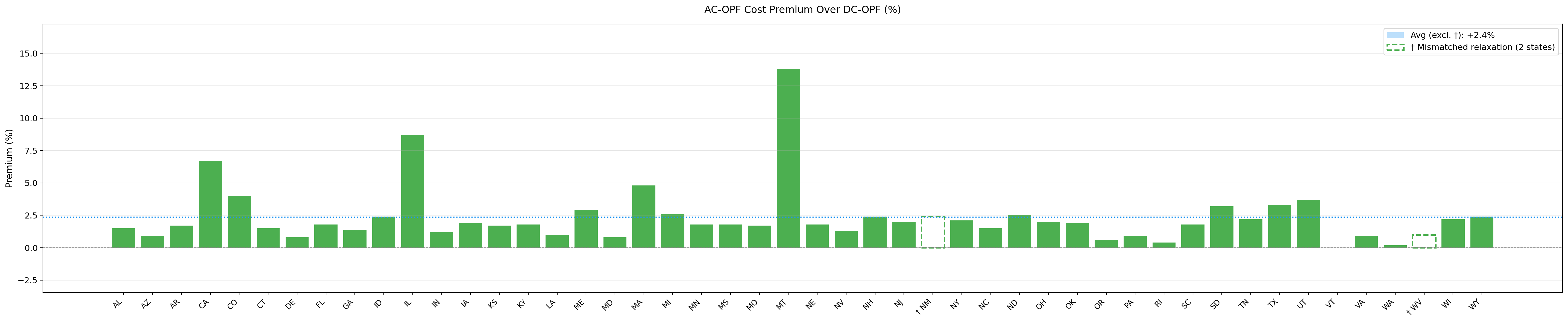}
\caption{AC-OPF cost premium over DC-OPF (\%) for each state.
  The average premium across states with matched relaxation levels
  is~+2.4\%.  New Mexico and West Virginia~(DC at~L0, AC at~L2)
  have mismatched relaxation levels, making their premiums not
  directly comparable.}
\label{fig:ac-premium}
\end{figure}

\textbf{AC--DC premium.}  The AC-OPF cost exceeds DC-OPF by 0.0--13.8\%
for L0 states, with a median premium of~1.8\%.  This
premium reflects resistive losses and reactive-power constraints ignored
by the DC linearization.  States with long radial corridors (Montana:
13.8\%, Massachusetts: 4.8\%) show the largest gaps.
The median value is consistent with the 1--5\% range reported in the
power systems literature for well-characterized
networks~\cite{molzahn2019}, providing evidence that the OSM-derived
models produce electrically plausible results despite estimated
parameters.
\Cref{fig:ac-premium} plots the premium for every state; states whose
DC and AC solves converged at different relaxation levels are shown
with dashed outlines and excluded from the average.

\begin{figure}[p]
\centering
\includegraphics[width=\textwidth]{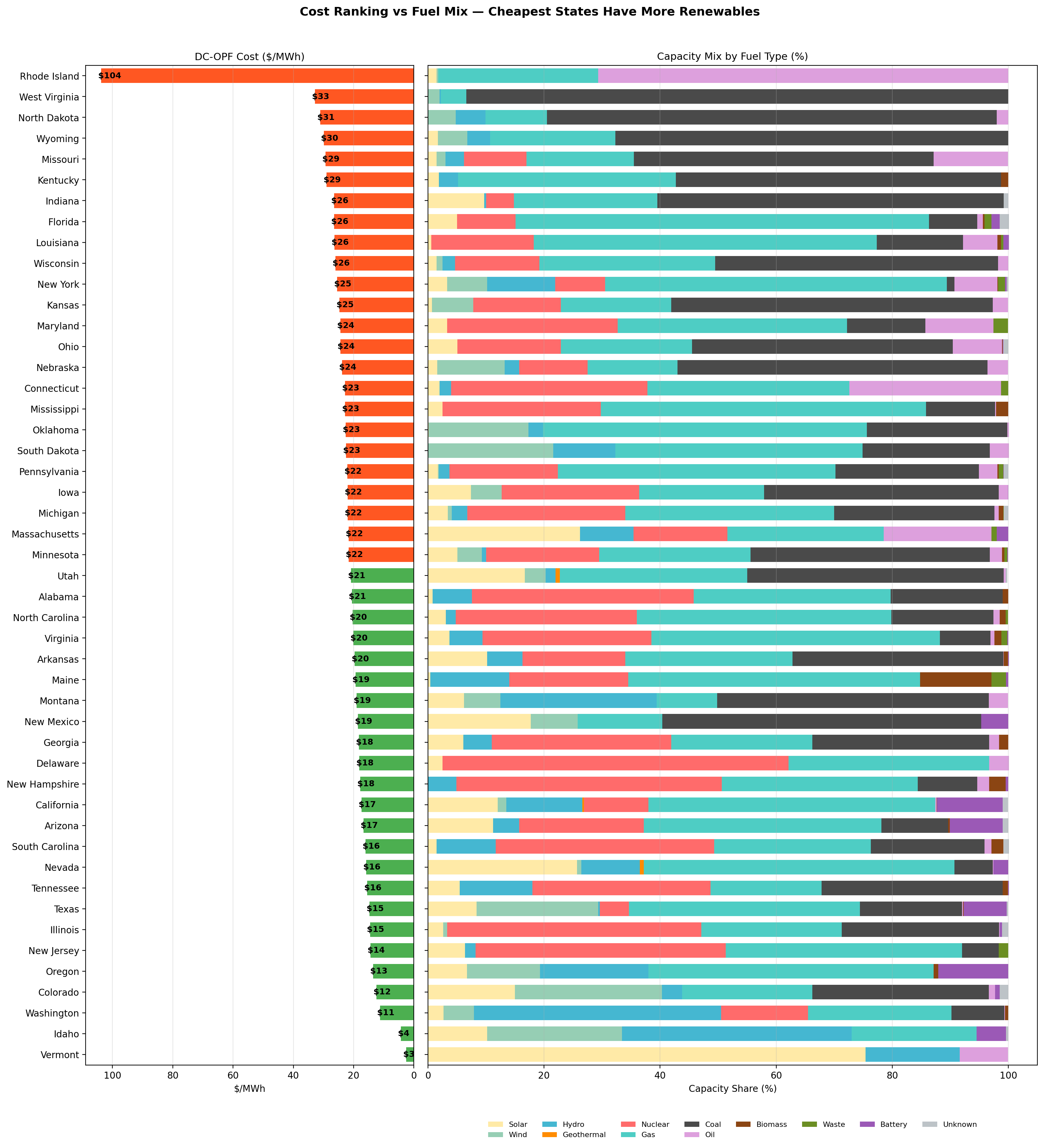}
\caption{States ranked by DC-OPF dispatch cost (\$/MWh, left panel)
  alongside their installed-capacity fuel mix (right panel).  States
  with high renewable or nuclear shares consistently achieve lower
  marginal costs.}
\label{fig:cost-fuelmix}
\end{figure}

\textbf{Cost vs.\ fuel mix.}  \Cref{fig:cost-fuelmix} ranks all
48~states by DC-OPF cost alongside their installed-capacity fuel mix.
States at the bottom of the ranking (lowest cost) have large shares of
hydro, wind, or nuclear capacity, all of which bid at near-zero
marginal cost.  States at the top tend to be dominated by gas and coal,
or have very small grids with limited generation options.
This ranking mirrors observed wholesale price patterns: EIA data show
that hydro-dominated states (e.g.\ Washington, Oregon) consistently
report the lowest average wholesale electricity rates, while
gas-dependent states rank among the
highest~\citep{eia_epa2024} -- confirming that the model's cost
structure responds correctly to fuel-mix composition.

\textbf{Losses.}  For the 42~L0 states, losses range from 0.2\%
(Washington) to 7.1\% (Montana), with a median of 1.0\%.
These values are physically plausible: real US transmission losses
average 2--3\%, and the model captures only high-voltage lines where
losses are relatively low.  Montana's elevated losses (7.1\%) stem from
long 230~kV corridors traversing sparsely populated terrain between
hydro sources and load centers.

\textbf{Solve time.}  AC-OPF solve time at the successful relaxation
level ranges from 17~seconds (Vermont, 40~buses) to 51~seconds
(Texas, 3{,}889~buses).  Models requiring elevated relaxation levels
spend additional time on failed attempts before reaching the successful
level; cumulative AC-OPF time (across all attempted relaxation
levels) for these states ranges from 1 to 8~minutes.

\subsection{Multi-State Regional Models}\label{sec:results-multi}

Six multi-state regional models were solved, ranging from a two-state
region to the full Eastern Interconnection (21{,}697~buses).  All six
achieve \texttt{LOCALLY\_SOLVED} for both DC-OPF and AC-OPF.
\Cref{tab:multi-state} summarizes the results.

\begin{table}[h!]
\centering\small
\caption{Multi-state regional AC-OPF results at 4~PM.}
\label{tab:multi-state}
\setlength{\tabcolsep}{3.5pt}
\begin{tabular}{@{}lcrrrrcrrrr@{}}
\toprule
Region & $N$ & Buses & Gens & Br & Load & Rlx &
  AC & DC & Loss & AC Time \\
 & & & & & (MW) & & (\$/MWh) & (\$/MWh) & (\%) & \\
\midrule
Pacific NW   &  2 & 1{,}106 &   132 &  1{,}935 &   3{,}364 & L0 &
  10.0 & 10.0 &  0.1 & 21\,s \\
New England  &  6 &    640 &   213 &  1{,}135 &   5{,}623 & L0 &
  16.7 & 16.7 &  0.2 & 19\,s \\
Desert SW    &  3 & 1{,}282 &   165 &  2{,}431 &  23{,}156 & L0 &
  18.3 & 18.1 &  0.6 & 25\,s \\
PJM          & 14 & 7{,}830 &   830 & 15{,}342 &  80{,}796 & L0 &
  20.5 & 20.5 &  0.6 & 2\,min \\
Western      & 11 & 5{,}076 &   746 &  9{,}511 &  80{,}754 & L0 &
  13.5 & 13.3 &  1.3 & 60\,s \\
Eastern      & 36 & 21{,}697 & 2{,}158 & 40{,}035 & 282{,}384 & L3 &
  17.9 & 17.8 &  0.5 & 47\,min \\
\bottomrule
\end{tabular}
\end{table}

\textbf{Pacific Northwest (OR\,+\,WA).}  This compact two-state region
produces the lowest dispatch cost (\$10.0/MWh), reflecting the region's
abundant hydropower.  The model solves at~L0 with only 0.1\% losses.

\textbf{New England (CT, MA, ME, NH, RI, VT).}  The six-state ISO-NE
footprint solves at~L0 with a cost of~\$16.7/MWh.  Notably, the merged
model eliminates the extreme \$104.1/MWh cost seen for Rhode Island in
isolation, because the merge provides import paths from neighbouring
Connecticut and Massachusetts generators.

\textbf{Desert Southwest (AZ, NV, UT).}  This three-state model
solves at~L0 (compared to Utah's L3 in isolation), confirming that
merging provides alternative transmission paths.  The marginal cost is
\$18.3/MWh.

\textbf{PJM (14~states, DC--VA).}  The PJM footprint is the largest L0
model (7{,}830~buses, 830~generators) and solves in 2~minutes of
AC-OPF time.  At \$20.5/MWh, the cost is within the range of real PJM
day-ahead LMPs, indicating plausibility rather than calibration.
Losses of 0.6\% are typical for a dense Eastern grid.

\textbf{Western Interconnection (11~states, AZ--WY).}  The full WECC
footprint (5{,}076~buses, 746~generators, 9{,}511~branches) solves
at~L0 in 60~seconds of solve time.  Losses of
1.3\% (1{,}066~MW) are physically plausible for a region spanning
11~states.

\textbf{Eastern Interconnection (36~states).}  The pipeline's largest
model at 21{,}697~buses and 2{,}158~generators solves at~L3 in
47~minutes.  The cost of \$17.9/MWh is lower than PJM's \$20.5,
reflecting the inclusion of low-cost hydro and wind states.  Losses of
0.5\% are lower than the Western model because
the Eastern grid has denser interconnections and shorter average
transmission distances.

\subsection{Peak vs.\ Off-Peak Comparison}\label{sec:results-temporal}



Each model was also solved at 4~AM (off-peak).
\Cref{tab:peak-offpeak} compares peak and off-peak results for the
largest models.
\begin{table}[h!]
\centering\small
\caption{Peak (4~PM) vs.\ off-peak (4~AM) comparison.}
\label{tab:peak-offpeak}
\begin{tabular}{@{}lrrcrrr@{}}
\toprule
Model & \multicolumn{3}{c}{4~PM (peak)} &
  \multicolumn{3}{c}{4~AM (off-peak)} \\
\cmidrule(lr){2-4}\cmidrule(lr){5-7}
 & Load & \$/MWh & Rlx & Load & \$/MWh & Rlx \\
 & (MW) & & & (MW) & & \\
\midrule
Texas       &  74{,}049 & 15.2 & L0 &  47{,}608 & 21.4 & L0 \\
California  &  31{,}806 & 18.5 & AC1 &  25{,}232 & 16.6 & L2 \\
Florida     &  27{,}522 & 26.9 & L0 &  15{,}789 & 25.3 & L0 \\
New York    &  27{,}746 & 26.0 & L5 &  19{,}115 & 30.3 & L3 \\
PJM         &  80{,}796 & 20.5 & L0 &  50{,}414 & 20.0 & L0 \\
Eastern     & 282{,}384 & 17.9 & L3 & 180{,}077 & 23.1 & L2 \\
Western     &  80{,}754 & 13.5 & L0 &  57{,}836 & 15.3 & L0 \\
\bottomrule
\end{tabular}
\end{table}

Loads drop by 21--43\% from peak to off-peak, consistent with typical
diurnal demand curves.  Across all 48~single-state models, 34~states
show higher per-MWh costs off-peak (median increase~\$2.7/MWh),
reflecting minimum-output constraints: at night, must-run
generators (nuclear, large coal units with high minimum-output levels)
form a larger fraction of total generation, reducing the system's ability
to dispatch cheaper units.  Solar output also drops to zero at 4~AM,
removing a zero-marginal-cost resource.  The remaining 14~states are
cheaper off-peak; among the table models, California
(\$18.5 $\to$ \$16.6), Florida (\$26.9 $\to$ \$25.3), and PJM
(\$20.5 $\to$ \$20.0) show this pattern, where night-time wind
production or reduced congestion offsets the must-run premium.

Relaxation levels are generally stable across hours.  Of the
48~single-state models, 44~(92\%) converge at~L0 off-peak, up from
42~(88\%) at peak.  Two states that required elevated relaxation at
peak improve to~L0 off-peak (New Mexico~L2$\to$L0,
West Virginia~L2$\to$L0); Illinois (L5$\to$L2), New York (L5$\to$L3),
and Utah (L3$\to$L2) also improve but remain above~L0.
California is the exception, shifting from AC1 (peak) to~L2
(off-peak) -- the lower demand reduces reactive stress but exposes
thermal constraints that the AC1 layer does not address.
Among the regional models, New York improves from L5 to~L3, and the
Eastern model from L3 to~L2, suggesting that lower demand partially
alleviates topology bottlenecks.  The Western model solves at~L0 for
both hours.

\section{Discussion}\label{sec:discussion}

\subsection{Limitations}

\textbf{Topology completeness.}  OSM typically captures one circuit per
transmission corridor; real networks may have 2--50 parallel circuits.
The topology and capacity factors (\cref{sec:topology-factors})
compensate for this under-representation, but they are calibrated
heuristics rather than measured data.  States where these factors
prove insufficient (Illinois, New York) require elevated
relaxation levels, and their solutions include load shedding that
reduces physical fidelity.

\textbf{Parameter accuracy.}  Line impedances are drawn from
voltage-class lookup tables, not utility engineering records.
Conductor type, spacing, bundling, and temperature are assumed from
typical practice.  Generator costs rely on EIA-923 heat-rate
data (available for $\sim$31\% of generators across all states) with
a fuel-type LUT fallback for the remainder; a further 80\% are
matched to EIA-860 for validated capacity and fuel type.  These approximations are sufficient for OPF convergence
but preclude precise congestion pricing or stability analysis.

\textbf{Demand model simplicity.}  Census-population-weighted
allocation is a reasonable spatial proxy, but real load patterns
depend on industrial facilities, commercial density, and weather.
The model uses a single hourly snapshot rather than a time series,
omitting storage cycling, ramp constraints, and dynamic stability.

\subsection{Comparison to Existing Approaches}

Compared to \emph{synthetic grids} such as the TAMU test
cases~\cite{birchfield2017}, our models preserve real geographic
correspondence: every bus and branch maps to a physical OSM feature
with coordinates.  Synthetic grids reproduce aggregate statistics
(degree distribution, impedance profiles) but cannot represent
actual transmission corridors or generation sites.

Compared to \emph{GridKit}~\cite{wiegmans2016}, which also extracts
topology from OSM, our pipeline goes substantially further:
it produces OPF-solvable models with calibrated impedances, thermal
ratings, generator cost curves, hourly demand allocation, and a
built-in solver.  GridKit provides topology only.

Compared to \emph{PyPSA-Eur}~\cite{horsch2018}, our work fills an
analogous role for the United States, where no ENTSO-E equivalent
provides standardized network data.  The challenges differ: US~OSM
coverage is less uniform than in Europe, and the lack of a central
grid operator necessitates the multi-source data-fusion approach
(EIA, HIFLD, Census) described in \cref{sec:demand}.

Critically, unlike prior OSM-based work that releases only code or
topology, we publicly release the complete solved models --
48~single-state and 6~multi-state regional networks with calibrated
parameters, demand profiles, and OPF solutions.

\section{Conclusion}\label{sec:conclusion}

We have demonstrated that usable transmission grid models can be
constructed entirely from open data.  The five-stage pipeline -- data
extraction, topology reconstruction, parameter estimation, demand
allocation, and optimal power flow -- transforms crowdsourced OSM
geometry and public EIA records into bus-branch models that converge
under AC-OPF for all 48~contiguous US states and multi-state regions
up to the full Eastern Interconnection.

The proposed pipeline yields models where 88\% of states
solve at the strictest constraint level.  Progressive relaxation
handles the remaining cases while transparently reporting which
constraints are binding, providing a built-in quality metric.

While the pipeline architecture
is general, the demand, generator, and Balancing Authority modules
rely on US-specific sources (EIA, HIFLD, Census Bureau).  Adapting
these modules for other countries with open EIA-equivalent data
(e.g.\ the ENTSO-E Transparency Platform for Europe) is a natural
extension.

We hope this work lowers the
barrier to transmission-level power systems research, enabling
students, policymakers, and researchers to study grid behavior
without requiring access to restricted proprietary data.  All
54~models are publicly available at
\url{https://github.com/microsoft/GridSFM}.  While the pipeline architecture is US-focused, the
methodology generalizes to any region with adequate OSM coverage and
public demand statistics.

\appendix

\section{Line Parameter Lookup Tables}\label{app:lut}

\Cref{tab:ac-overhead} lists the per-kilometer parameters for AC
overhead lines.  \Cref{tab:cable} lists underground cable parameters.
\Cref{tab:xfmr-lut} lists representative transformer parameters for the
most common voltage pairs (9 of 52 entries shown; the full table is
voltage-pair-specific).  All resistance values assume
75\textdegree{}C conductor temperature.

\begin{table}[h!]
\centering\small
\caption{AC overhead line parameters (conservative mode).}
\label{tab:ac-overhead}
\begin{tabular}{@{}rrrrrll@{}}
\toprule
kV & $R$ & $X$ & $B$ & MVA & Conductor & Bundle \\
   & ($\Omega$/km) & ($\Omega$/km) & ($\mu$S/km) & & & \\
\midrule
765  & 0.0076 & 0.267 & 5.46 & 2400 & 4$\times$Bluebird 2156\,kcmil & 12\,in \\
525  & 0.0100 & 0.290 & 4.60 & 2000 & 3$\times$Bluebird 2156\,kcmil &  -  \\
345  & 0.0200 & 0.370 & 3.50 & 1000 & 2$\times$Bluebird 2156\,kcmil &  -  \\
230  & 0.0280 & 0.450 & 3.00 &  600 & 1$\times$Drake 795\,kcmil &  -  \\
161  & 0.0350 & 0.460 & 2.80 &  400 & 1$\times$Drake (TVA/MISO) &  -  \\
138  & 0.0400 & 0.450 & 2.60 &  300 & 1$\times$Drake 795\,kcmil &  -  \\
115  & 0.0450 & 0.450 & 2.40 &  250 & 1$\times$Dove 556.5\,kcmil &  -  \\
 69  & 0.0600 & 0.470 & 2.00 &  150 & 1$\times$Partridge 266.8\,kcmil &  -  \\
\bottomrule
\end{tabular}
\end{table}

\begin{table}[h!]
\centering\small
\caption{Underground cable parameters (XLPE insulation, conservative
  mode).}
\label{tab:cable}
\begin{tabular}{@{}rrrrr@{}}
\toprule
kV & $R$ ($\Omega$/km) & $X$ ($\Omega$/km) & $B$ ($\mu$S/km) & MVA \\
\midrule
500 & 0.0130 & 0.125 & 50.0 & 1400 \\
230 & 0.0250 & 0.115 & 30.0 &  500 \\
138 & 0.0350 & 0.105 & 20.0 &  250 \\
 69 & 0.0550 & 0.090 & 12.0 &  120 \\
\bottomrule
\end{tabular}
\end{table}

\begin{table}[h!]
\centering\small
\caption{Representative transformer parameters by voltage pair.}
\label{tab:xfmr-lut}
\begin{tabular}{@{}llrrrr@{}}
\toprule
HV (kV) & LV (kV) & $X$ (pu) & $R$ (pu) & MVA & Notes \\
\midrule
765 & 500 & 0.10 & 0.002 & 1500 & Autotransformer \\
765 & 345 & 0.12 & 0.002 & 1200 & Large ratio \\
500 & 345 & 0.08 & 0.002 & 1200 & Standard EHV \\
345 & 230 & 0.08 & 0.003 &  800 & EHV/HV boundary \\
230 & 138 & 0.09 & 0.004 &  400 & HV/subtransmission \\
230 &  69 & 0.10 & 0.005 &  250 & Large step-down \\
138 &  69 & 0.08 & 0.005 &  150 & Common subtransmission \\
115 &  69 & 0.07 & 0.005 &  150 & Subtransmission \\
 69 & 34.5 & 0.07 & 0.006 &   50 & Subtrans/distribution \\
\bottomrule
\end{tabular}
\end{table}

\section{Generator Parameter Lookup Tables}\label{app:gen-params}

\Cref{tab:gen-costs} lists the default marginal cost and operational
parameters by fuel type; \cref{tab:heat-rates} lists the heat-rate
defaults used when no EIA-923 match is available.

\begin{table}[h!]
\centering\small
\caption{Generator cost and operational defaults by fuel type.}
\label{tab:gen-costs}
\begin{tabular}{@{}lrrrrr@{}}
\toprule
Fuel & $c_1$ & $c_0$ & Startup & $P_\text{min}$ & PF \\
     & (\$/MWh) & (\$/h) & (\$) & (\%) & \\
\midrule
Nuclear     &  12.0 & 100 & 50{,}000 & 50 & 0.90 \\
Coal        &  35.0 &  50 & 10{,}000 & 30 & 0.85 \\
Gas (CCGT)  &  26.0 &  20 &  2{,}000 & 20 & 0.85 \\
Gas turbine &  70.0 &  10 &      500 &  0 & 0.85 \\
Oil         &  80.0 &  30 &  1{,}000 & 10 & 0.85 \\
Diesel      &  90.0 &  20 &      500 &  0 & 0.85 \\
Biomass     &  45.0 &  30 &  3{,}000 & 20 & 0.85 \\
Waste       &  40.0 &  40 &  5{,}000 & 30 & 0.85 \\
Geothermal  &   5.0 &  50 &  1{,}000 & 70 & 0.85 \\
Hydro       &   8.0 &   0 &        0 &  0 & 0.80 \\
Solar       &   0.0 &   0 &        0 &  0 & 0.95 \\
Wind        &   0.0 &   0 &        0 &  0 & 0.95 \\
Battery     &  15.0 &   0 &        0 &  0 & 0.95 \\
\bottomrule
\end{tabular}
\end{table}

\begin{table}[h!]
\centering\small
\caption{Heat-rate defaults for thermal generators (used when no
  EIA-923 plant match is available).}
\label{tab:heat-rates}
\begin{tabular}{@{}lrrr@{}}
\toprule
Fuel & Heat rate & Fuel price & VOM \\
     & (BTU/kWh) & (\$/MMBtu) & (\$/MWh) \\
\midrule
Nuclear     & 10{,}400 &  0.80 & 2.0 \\
Coal        &  9{,}800 &  2.50 & 5.0 \\
Gas (CCGT)  &  6{,}600 &  3.50 & 3.0 \\
Gas turbine & 10{,}000 &  3.50 & 4.0 \\
Oil         & 10{,}500 & 15.00 & 4.0 \\
Diesel      &  9{,}500 & 20.00 & 5.0 \\
Biomass     & 12{,}000 &  2.00 & 8.0 \\
Waste       & 14{,}000 &  0.00 & 15.0 \\
\bottomrule
\end{tabular}
\end{table}

\section{Per-Unit Conversion}\label{app:perunit}

It is convenient to represent quantities used in the model to avoid numerical values spanning many orders of magnitudes. This is done by defining some ``base'' units and normalizing other quantities with respect to them. The resulting numbers are technically unit-less and they are referred to as per unit values. In our pipeline, we adopt the following base and per unit quantities:
\begin{align}
  S_\text{base} &= 100~\text{MVA} \\
  V_\text{base} &= \text{nominal bus voltage (kV)} \\
  Z_\text{base} &= V_\text{base}^2 / S_\text{base} \\
  Z_\text{pu}   &= Z / Z_\text{base}
  \qquad P_\text{pu} = P_\text{MW} / S_\text{base}
\end{align}
The pipeline performs SI-to-per-unit conversion at the boundary between
Steps~2--4 (which operate in physical units) and Step~5 (which expects
per-unit input).  Branch impedances, shunt admittances, generator
limits, and load values are all converted during the export step.

\bibliography{references}

@article{arderne2020,
  author  = {Arderne, Christopher and Zorn, Conrad and Nicolas, Claire and Koks, Elco E.},
  title   = {Predictive mapping of the global power system using open data},
  journal = {Scientific Data},
  volume  = {7},
  pages   = {19},
  year    = {2020},
  doi = {10.1038/s41597-019-0347-4}
}

@article{birchfield2017,
  author  = {Birchfield, Adam B. and Xu, Ti and Gegner, Kathleen M. and Shetye, Komal S. and Overbye, Thomas J.},
  title   = {Grid Structural Characteristics as Validation Criteria for Synthetic Networks},
  journal = {IEEE Transactions on Power Systems},
  volume  = {32},
  number  = {4},
  pages   = {3258--3265},
  year    = {2017},
  doi = {10.1109/TPWRS.2016.2616385}
}

@inproceedings{coffrin2018,
  author    = {Coffrin, Carleton and Bent, Russell and Sundar, Kaarthik and Ng, Yeesian and Lubin, Miles},
  title     = {{PowerModels.jl}: An Open-Source Framework for Exploring Power Flow Formulations},
  booktitle = {Proceedings of the Power Systems Computation Conference (PSCC)},
  year      = {2018},
}

@article{horsch2018,
  author  = {H{\"o}rsch, Jonas and Hofmann, Fabian and Schlachtberger, David and Brown, Tom},
  title   = {{PyPSA-Eur}: An Open Optimisation Model of the {European} Transmission System},
  journal = {Energy Strategy Reviews},
  volume  = {22},
  pages   = {207--215},
  year    = {2018},
  doi     = {10.1016/j.esr.2018.08.012}
}

@article{molzahn2019,
  author  = {Molzahn, Daniel K. and Hiskens, Ian A.},
  title   = {A Survey of Relaxations and Approximations of the Power Flow Equations},
  journal = {Foundations and Trends in Electric Energy Systems},
  volume  = {4},
  number  = {1--2},
  pages   = {1--221},
  year    = {2019},
}

@misc{nerc2023,
  author = {{North American Electric Reliability Corporation}},
  title  = {Critical Infrastructure Protection Standards {CIP-002} through {CIP-014}},
  year   = {2023},
  url    = {https://www.nerc.com/standards/reliability-standards/cip},
}

@article{wachter2006,
  author  = {W{\"a}chter, Andreas and Biegler, Lorenz T.},
  title   = {On the Implementation of an Interior-Point Filter Line-Search Algorithm for Large-Scale Nonlinear Programming},
  journal = {Mathematical Programming},
  volume  = {106},
  pages   = {25--57},
  year    = {2006},
  doi = {10.1007/s10107-004-0559-y}
}

@misc{wiegmans2016,
  author = {Wiegmans, Bart},
  title  = {{GridKit}: {European} and {North American} Extracts},
  year   = {2016},
  url    = {https://zenodo.org/record/47317},
  note   = {Dataset, Zenodo, accessed 2026},
}

@misc{ferc2023,
  author = {{FERC}},
  title  = {Critical Energy Infrastructure Information ({CEII})},
  year   = {2023},
  note   = {18 CFR \S~388.113},
  url    = {https://www.ecfr.gov/current/title-18/chapter-I/subchapter-X/part-388/section-388.113},
}

@misc{osm2026,
  author = {{OpenStreetMap Contributors}},
  title  = {{OpenStreetMap}},
  year   = {2026},
  url    = {https://www.openstreetmap.org},
  note = {Accessed 2026}
}

@misc{eia930_2026,
  author = {{U.S. Energy Information Administration}},
  title  = {{EIA-930} Hourly Electric Grid Monitor},
  year   = {2026},
  url    = {https://www.eia.gov/electricity/gridmonitor/},
  note = {Accessed 2026}
}

@misc{eia860_2024,
  author = {{U.S. Energy Information Administration}},
  title  = {Form {EIA-860}: Annual Electric Generator Report},
  year   = {2024},
  url    = {https://www.eia.gov/electricity/data/eia860/},
  note = {Accessed 2026}
}

@misc{eia923_2024,
  author = {{U.S. Energy Information Administration}},
  title  = {Form {EIA-923}: Power Plant Operations Report},
  year   = {2024},
  url    = {https://www.eia.gov/electricity/data/eia923/},
  note = {Accessed 2026}
}

@misc{census2024,
  author = {{U.S. Census Bureau}},
  title  = {American Community Survey 5-Year Estimates},
  year   = {2024},
  url    = {https://data.census.gov},
  note = {Accessed 2026}
}

@misc{pglib2019,
      title={The Power Grid Library for Benchmarking AC Optimal Power Flow Algorithms}, 
      author={Sogol Babaeinejadsarookolaee and Adam Birchfield and Richard D. Christie and Carleton Coffrin and Christopher DeMarco and Ruisheng Diao and Michael Ferris and Stephane Fliscounakis and Scott Greene and Renke Huang and Cedric Josz and Roman Korab and Bernard Lesieutre and Jean Maeght and Terrence W. K. Mak and Daniel K. Molzahn and Thomas J. Overbye and Patrick Panciatici and Byungkwon Park and Jonathan Snodgrass and Ahmad Tbaileh and Pascal Van Hentenryck and Ray Zimmerman},
      year={2021},
      eprint={1908.02788},
      archivePrefix={arXiv},
      primaryClass={math.OC},
      url={https://arxiv.org/abs/1908.02788}, 
}

@techreport{utaustin2021,
  author      = {King, Carey W. and Kutanoglu, Erhan and Leibowicz, Benjamin D. and Lin, Ning and Niyogi, Dev and Rai, Varun and Rhodes, Joshua D. and Santoso, Surya and Spence, David and Tompaidis, Stathis and Zarnikau, Jay and Zhu, Hao},
  title       = {The Timeline and Events of the {February} 2021 {Texas} Electric Grid Blackouts},
  institution = {The University of Texas at Austin Energy Institute},
  year        = {2021},
  url         = {https://energy.utexas.edu/research/ercot-blackout-2021},
  note = {Accessed 2026}
}

@misc{climatecentral2024,
  author = {{Climate Central}},
  title  = {Weather-Related Power Outages Rising},
  year   = {2024},
  url    = {https://www.climatecentral.org/climate-matters/weather-related-power-outages-rising},
  note = {Accessed 2026}
}

@techreport{iea_energy_ai2025,
  author       = {{International Energy Agency}},
  title        = {Energy and AI},
  year         = {2025},
  institution  = {IEA},
  url          = {https://www.iea.org/reports/energy-and-ai/},
  note         = {Published 10 April 2025}
}

@inproceedings{xu2017irep,
  author       = {Xu, Ti and Birchfield, Adam B. and Shetye, Komal S. and Overbye, Thomas J.},
  title        = {Creation of Synthetic Electric Grid Models for Transient Stability Studies},
  booktitle = {Proceedings of the IREP Symposium (Bulk Power System Dynamics and Control)},
  year         = {2017},
  url          = {https://overbye.engr.tamu.edu/wp-content/uploads/sites/146/2022/01/IREP_Ti_WithFooter_ARCHIVE.pdf}
}

@misc{osm_stats,
  author       = {{OpenStreetMap Wiki contributors}},
  title        = {Stats --- OpenStreetMap Wiki},
  year         = {2026},
  url          = {https://wiki.openstreetmap.org/wiki/Stats},
  note         = {Accessed 2026}
}

@misc{osm_taginfo,
  author       = {Topf, Jochen and contributors},
  title        = {OpenStreetMap Taginfo},
  year         = {2026},
  url          = {https://taginfo.openstreetmap.org/},
  note         = {Tag usage statistics, accessed 2026-03-30}
}

@techreport{eia_epa2024,
  author       = {{U.S. Energy Information Administration}},
  title = {Electric Power Annual 2024},
  year         = {2025},
  institution  = {U.S. EIA},
  url          = {https://www.eia.gov/electricity/annual/},
  note  = {Published 2025, data year 2024, accessed 2026},
}

@misc{hifld_ba,
  author = {{U.S. Department of Homeland Security}},
  title  = {Electric Planning Areas (Balancing Authorities)},
  year   = {2025},
  url    = {https://services5.arcgis.com/HDRa0B57OVrv2E1q/arcgis/rest/services/Electric_Planning_Areas/FeatureServer/0},
  note   = {Homeland Infrastructure Foundation-Level Data (HIFLD); ArcGIS Feature Server},
  note = {Accessed 2026}
}

@book{kirschen2024power,
  title={Power Systems: Fundamental Concepts and the Transition to Sustainability},
  author={Kirschen, Daniel S},
  year={2024},
  publisher={John Wiley \& Sons}
}

@book{glover2012power,
  title={Power system analysis and design},
  author={Glover, J Duncan and Sarma, Mulukutla S and Overbye, Thomas Jeffrey and Padhy, NP},
  volume={2008},
  year={2012},
  publisher={Cengage Learning Stamford, CT, USA}
}

@book{Low_draft,
title={Power System Analysis: Analytical tools and structural properties},
year={2026},
author={Steven Low},
publisher={Cambridge University Press}
}

@book{bergen2009power,
  title={Power systems analysis},
  author={Bergen, Arthur R},
  year={2009},
  publisher={Pearson Education India}
}

@book{kirschen2026fundamentals,
  title={Fundamentals of power system economics},
  author={Kirschen, Daniel S and Strbac, Goran},
  year={2026},
  publisher={John Wiley \& Sons}
}

@misc{eia861_2024,
  author = {{U.S. Energy Information Administration}},
  title  = {Form {EIA-861}: Annual Electric Power Industry Report},
  year   = {2024},
  url    = {https://www.eia.gov/electricity/data/eia861/},
  note    = {Accessed 2026},
}

@article{zhang2012geometry,
  title={Geometry of injection regions of power networks},
  author={Zhang, Baosen and Tse, David},
  journal={IEEE Transactions on Power Systems},
  volume={28},
  number={2},
  pages={788--797},
  year={2012},
  publisher={IEEE}
}

@article{low2014convex1,
  title={Convex relaxation of optimal power flow--Part I: Formulations and equivalence},
  author={Low, Steven H},
  journal={IEEE Transactions on Control of Network Systems},
  volume={1},
  number={1},
  pages={15--27},
  year={2014},
  publisher={IEEE}
}

@article{low2014convex2,
  title={Convex relaxation of optimal power flow—Part II: Exactness},
  author={Low, Steven H},
  journal={IEEE Transactions on Control of Network Systems},
  volume={1},
  number={2},
  pages={177--189},
  year={2014},
  publisher={IEEE}
}

@article{taylor1975demand,
  title={The demand for electricity: a survey},
  author={Taylor, Lester D},
  journal={The Bell Journal of Economics},
  pages={74--110},
  year={1975},
  publisher={JSTOR}
}

@article{hyndman2009density,
  title={Density forecasting for long-term peak electricity demand},
  author={Hyndman, Rob J and Fan, Shu},
  journal={IEEE Transactions on Power Systems},
  volume={25},
  number={2},
  pages={1142--1153},
  year={2009},
  publisher={IEEE}
}

\end{document}